\documentclass[aps,prd,letterpaper,showpacs,showkeys,twocolumn,floatfix,nofootinbib,superscriptaddress]{revtex4-1} 

\usepackage{amsmath,amssymb,amsfonts}
\usepackage{bm}
\usepackage{graphicx}
\usepackage{subfigure}

\usepackage[svgnames]{xcolor}

\begin{document}

\preprint{ADP-12-13/T780}

\title{Dihadron Fragmentation Functions from the NJL-jet model and their QCD Evolution}

\author{Andrew~Casey}
\affiliation{CSSM and ARC Centre of Excellence for Particle Physics at the Terascale,\\ 
School of Chemistry and Physics, \\
University of Adelaide, Adelaide SA 5005, Australia
\\ http://www.physics.adelaide.edu.au/cssm
}

\author{Ian~C.~Clo\"et}
\affiliation{CSSM and ARC Centre of Excellence for Particle Physics at the Terascale,\\ 
School of Chemistry and Physics, \\
University of Adelaide, Adelaide SA 5005, Australia
\\ http://www.physics.adelaide.edu.au/cssm
}

\author{Hrayr~H.~Matevosyan}
\affiliation{CSSM and ARC Centre of Excellence for Particle Physics at the Terascale,\\ 
School of Chemistry and Physics, \\
University of Adelaide, Adelaide SA 5005, Australia
\\ http://www.physics.adelaide.edu.au/cssm
}

\author{Anthony~W.~Thomas}
\affiliation{CSSM and ARC Centre of Excellence for Particle Physics at the Terascale,\\ 
School of Chemistry and Physics, \\
University of Adelaide, Adelaide SA 5005, Australia
\\ http://www.physics.adelaide.edu.au/cssm
}

\begin{abstract}
We present results for dihadron fragmentation functions from the NJL-jet model evolved from the model scale to a typical experimental scale of $4~\mathrm{GeV}^2$. The numerical method used in this evolution is discussed in detail. The effect of evolution on the shapes of the dihadron fragmentation functions is discussed for a doubly favored process~($u\to\pi^+\pi^-$), as well as a singly favored~($u\to\pi^+K^-$) process involving light quarks. Finally, we explore the production of $K^+ K^-$ pairs from an initial $u$, $d$ or $s$ quark.
\end{abstract}

\maketitle

\section{Introduction}

Experimental processes such as deep-inelastic scattering~(DIS), semi-inclusive deep-inelastic scattering~(SIDIS) and Drell-Yan~(DY) have provided invaluable information about the structure of the nucleon~\cite{PhysRevD.59.014033,PhysRevLett.85.2892,Thomas:2001kw,Beckmann:2002rr,Airapetian:2003ct,Barone:2003jp,Airapetian:2004zf,Airapetian:2007mh,Cloet:2009qs,Alekseev:2010ub,Gross:2012sj}. With several new experimental facilities with 100\% duty factor under construction, SIDIS will play an increasingly important role in the development of our theoretical and experimental understanding of the structure of the nucleon. The elusive $s-\bar{s}$ asymmetry~\cite{Signal:1987gz,Barone:1999yv,Davidson:2001ji,Bentz:2009yy} is one area of interest that may finally be pinned down through the results obtained at these new facilities. The distribution of the spin of the proton~\cite{Ji:1996ek,Ji:1996nm,Bratt:2010jn,QCDSF:2011aa,Hagler:2011zzb,Thomas:2010zzc,Thomas:2008ga,Myhrer:2007cf,Bass:2009ed,Bass:2011zn,Wakamatsu:2009gx,Wakamatsu:2010qj,Wakamatsu:2010cb,:2008px,Bacchetta:2008xw,Barone:2001sp,Leader:2010gz,Leader:2010rb,Leader:2011za,Leader:2011zz,Vogelsang:2009zz} is an area of current excitement where polarized SIDIS is potentially extremely valuable through the study of tranverse momentum dependent parton distribution functions~\cite{Ceccopieri:2006wy,Bacchetta:2006tn,Cloet:2007em,Gamberg:2010uw,Gamberg:2010xi,Wollny:2010zp,Anselmino:2011ay,Anselmino:2011ch,Aybat:2011ta,Gamberg:2011my,Bacchetta:2011gx}, which will complement work on generalized parton distributions~\cite{Goeke:2001tz,Burkardt:2002hr,Diehl:2003ny,Ji:2004gf,Belitsky:2005qn,Boffi:2007yc,Hagler:2011zzb}.

To allow these studies to fulfil their potential, we must develop a deep understanding of the fragmentation functions~\cite{Radici:2011ix}, particularly their flavor, spin and transverse momentum dependence.  Fragmentation functions appear in certain scattering reactions, for 
example, in SIDIS experiments~\cite{Hirai:2007cx,deFlorian:2007aj} and in $e^+ e^-$ annihilation reactions \cite{Biebel:520981,deFlorian:2003cg,Hirai:2006au,Hirai:2007aa,Francisconi:2010zz}. Experiments are planned to use SIDIS to probe the flavor dependence of the parton distribution functions (PDFs), for example, and therefore understanding fragmentation functions has become very important. Phenomenological extraction of fragmentation functions suffers from significant uncertainty, even for favored fragmentation functions, which effects the systematic errors associated with extracting the flavor dependence of PDFs through SIDIS. The increasing interest in SIDIS experiments led to the development of the NJL-jet model~\cite{Ito:2009zc,Matevosyan:2010hh,Matevosyan:2011ey,Matevosyan:2011vj}, which builds on the Field-Feynman quark-jet model~(FFQJM)~\cite{Field:1977fa}, by using an effective chiral quark model to provide a unified framework in which calculations of both quark distribution and fragmentation functions can be performed. NJL-jet model calculations of pion fragmentation functions were obtained in Ref.~\cite{Ito:2009zc}. The NJL-jet model was extended to include strange quark contributions and kaon fragmentation functions were calculated in Ref.~\cite{Matevosyan:2010hh}. Further extensions of the model involved the inclusion of vector meson, nucleon and antinucleon fragmentation channels~\cite{Matevosyan:2011ey}, as well as the study of their transverse momentum dependence~\cite{Matevosyan:2011vj} and Collins fragmentation functions~\cite{Matevosyan:2012ga,Matevosyan:2012ms,Matevosyan:2012hocm}.

The probability of a fragmenting quark to produce two hadrons is represented by dihadron fragmentation functions~(DFFs). DFFs have been studied recently in Refs.~\cite{Bacchetta:2006un,Zhou:2011ba} in order to understand their dependence on invariant mass of the two produced hadrons. The focus of Ref.~\cite{Bacchetta:2006un} was to fit parameters for a spectator model to output from the PYTHIA event generator~\cite{Sjostrand:2000wi} tuned for HERMES experiments~\cite{Liebing:2004us} for DFFs with a dependence on the sum of the light-cone momentum fractions of the two produced hadrons and their invariant mass squared. Ref.~\cite{Zhou:2011ba} focused on studying DFFs for large invariant mass. DFFs with no invariant mass dependence were studied in the NJL-jet model in Ref.~\cite{PhysRevD.85.114049} at the model momentum scale of $Q^2_0=0.2~\mathrm{GeV}^2$. In order to compare the results with experimental data, we need to evolve the DFFs up to a typical experimental scale. The evolution equations for the DFFs are derived in Ref.~\cite{deFlorian:2003cg} from factorization of the cross-section for the production of two hadrons in $e^+e^-$ annihilation in the $\overline{\text{MS}}$ factorization scheme. In Ref.~\cite{Majumder:2004wh}, the non- quark evolution equations for DFFs were studied, while Ref.~\cite{Majumder:2004br} focused on the QCD evolution equations for singlet quark and gluon DFFs. The ratio of the dihadron and single hadron fragmentation functions, which is useful when considering experimental measurements, was also examined in Refs.~\cite{Majumder:2004wh,Majumder:2004br}. Initial conditions for DFFs for different pairs of hadrons and different values of $z_1$ and $z_2$ are investigated in Ref.~\cite{Grigoryan:2008ut}, with a focus on the correlation function $R_{cor}$ obtained in the FFQJM~\cite{Field:1977fa}.

An area of current interest in which the dihadron fragmentation functions of quarks may be useful are transversity distributions~\cite{Barone:2001sp}. Transversity distributions are one of the three leading-twist distribution functions that don't vanish when integrated over the transverse momentum. They describe the quark structure of the nucleon~(the other two being unpolarized and helicity quark distribution functions) and these functions enter into asymmetries with chiral-odd versions of a special type of DFF known as interference fragmentation functions~(IFFs)~\cite{Boffi:1999it,Radici:1999sc,Radici:2002zg,Bacchetta:2004it,She:2007ht}. IFFs are DFFs with a dependence on the polarization of the fragmenting quark. In Refs.~\cite{Bacchetta:2011ip,Courtoy:2011ni,PhysRevD.85.114023}, it was suggested that DFFs may be useful in extracting transversity distributions by considering the SIDIS production of two hadrons with small invariant mass. Transversity distribution functions are not a focus of this paper, but are presented as motivation for further investigation into DFFs.

This work focuses on performing QCD evolution of the DFFs from the NJL-jet model momentum scale of $Q^2_0=0.2~\mathrm{GeV}^2$ to a typical experimental momentum scale of $Q^2=4~\mathrm{GeV}^2$. In Section~\ref{sec:qsffdff} we present a brief summary of fragmentation function equations from which the model scale solutions were obtained and used as input for the evolution equations of the DFFs. Section~\ref{sec:evolsff} describes the method for solving the evolution equations for single hadron fragmentation functions~(SFFs), which are needed for the evolution of the DFFs. It also serves as a simple version of the method used to solve the DFF evolution equations, while the method for solving the evolution equations for the DFFs is described in Section~\ref{sec:evoldffs}. A comparison of the model scale and evolved scale DFFs is presented in Section~\ref{sec:evolresults}.  Section~\ref{sec:comparison} shows how the evolution code works on data from Ref.~\cite{Majumder:2004br} as well as comparing our solutions to that data. Our data is evolved to a range of values of $Q^2$ in this section to display how the up quark and gluon DFFs change for larger values of $Q^2$.
\\
\section{Single Hadron and Dihadron Fragmentation Functions from the NJL-jet model}
\label{sec:qsffdff}

In Ref.~\cite{PhysRevD.85.114049}, integral equations for the single hadron and dihadron fragmentation functions from the NJL-jet model are described, and the method employed to solve them at the model scale of $Q^2_0=0.2~\mathrm{GeV}^2$ is presented. SFFs appear in the cross section for SIDIS experiments and thus play an important part in the theoretical understanding of these experiments. In the NJL-jet model the SFFs, $D^h_q(z)$, which correspond to the probability of producing a hadron $h$ with light-cone momentum fraction $z$ from a fragmenting quark $q$, are given by~\cite{Ito:2009zc}
\begin{align}
D^h_q(z) & =\hat{d}^h_q(z)+\sum_{Q}\int^{1}_{z}\frac{dy}{y}\,\,\, \hat{d}^Q_q\left(\frac{z}{y}\right)D^h_Q(y).
\label{EQ_SINGEL_FRAG}
\end{align}
The first term on the right hand side of Eq.~\eqref{EQ_SINGEL_FRAG} is the renormalized elementary quark fragmentation function, which corresponds to the process where the detected hadron is the only emitted hadron. We refer to this term as the driving function. The second term corresponds to the probability of emitting a hadron after the first emission step in the quark cascade and these terms have a sizeable effect at low values of $z$, while vanishing for higher $z$ values. To solve the second term we use $\hat{d}^Q_q(z)  =\hat{d}^h_q(1-z)|_{h=q\bar{Q}}$ to write all functions in terms of their relation to the emitted hadron $h$.

Dihadron fragmentation functions are another important tool in the theoretical understanding of the structure of hadrons. In the NJL-jet model, the DFF are given by
\begin{widetext}
\begin{multline}
\label{DFFcv}
 D^{h_1,h_2}_q(z_1,z_2)  =  \hat{d}^{h_1}_q(z_1)\frac{D^{h_2}_{q_1}\left(\frac{z_2}{1-z_1}\right)}{1-z_1}  +\hat{d}^{h_2}_q(z_2)\frac{D^{h_1}_{q_2}\left(\frac{z_1}{1-z_2}\right)}{1-z_2} \\  +\sum_Q\int^{\frac{z_1}{z_1+z_2}}_{z_1}d\xi_1\int^{\frac{z_2}{z_1+z_2}}_{z_2}d\xi_2\delta(z_2\xi_1-z_1\xi_2)\hat{d}^Q_q(z_1/\xi_1)D^{h_1,h_2}_Q(\xi_1,\xi_2),
\end{multline}
\end{widetext}
where the first term corresponds to the probability of producing hadron $h_1$ from the quark $q$ at the first emission step in the cascade, followed by hadron $h_2$ produced either directly afterwards or further down in the quark decay chain, while the second term is similar to the first one, except for $h_1 \leftrightarrow h_2$ . These two terms constitute the driving function of the DFFs, similar to the first term in Eq.~\eqref{EQ_SINGEL_FRAG}. The third term on the right hand side of Eq.~\eqref{DFFcv} corresponds to the probability of having both the detected hadrons produced after the first hadron emission. DFFs correspond to the probability of producing two hadrons, $h_1$ and $h_2$, in the decay chain of a fragmenting quark $q$, with light-cone momentum fractions $z_1$ and $z_2$, respectively. 

\begin{figure}[t]
\begin{center}
\subfigure[]{\label{fig:upplpmi3dmodel}
\includegraphics[width=0.48\textwidth]{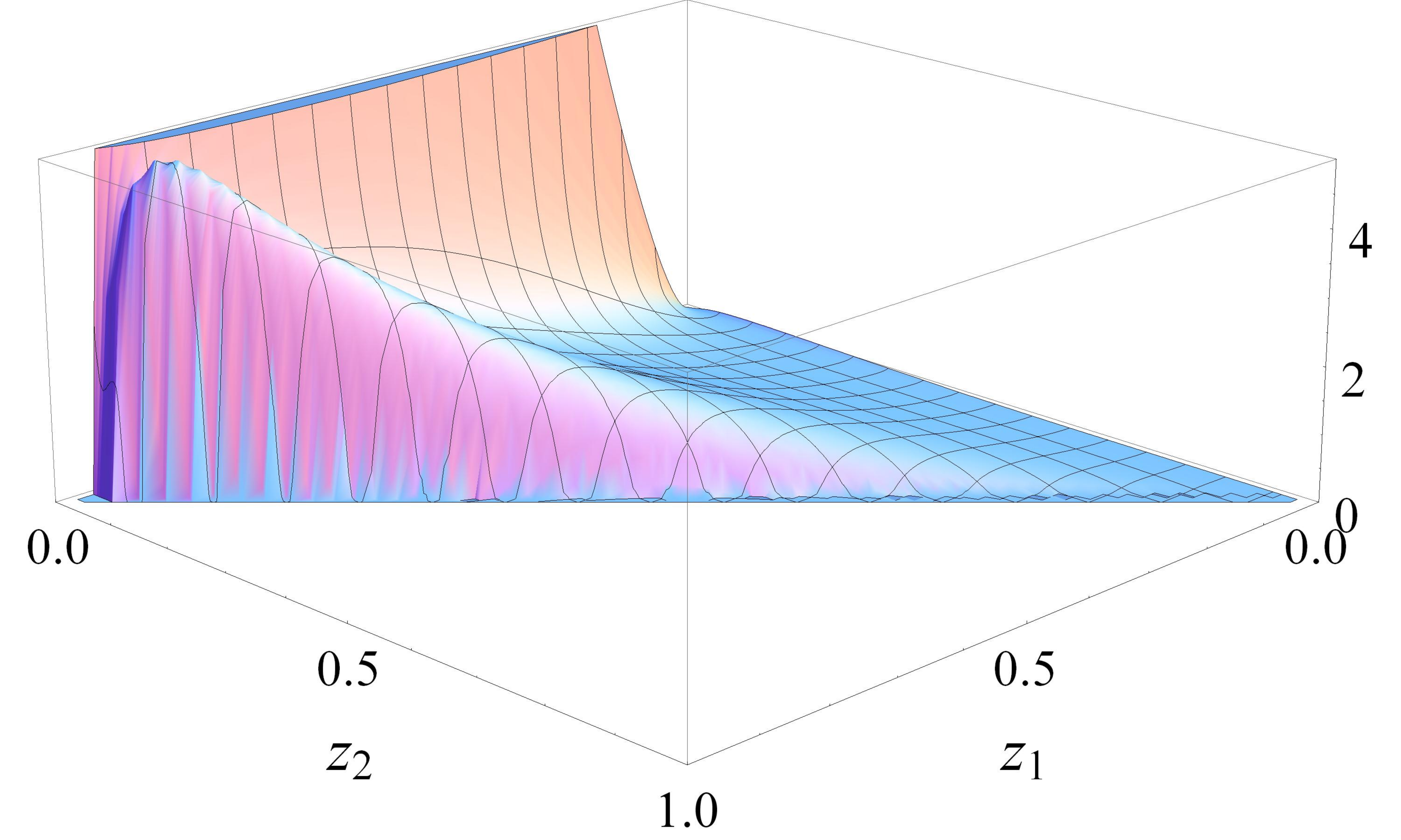}}
\subfigure[]{\label{fig:upplpmi3devol}
\includegraphics[width=0.48\textwidth]{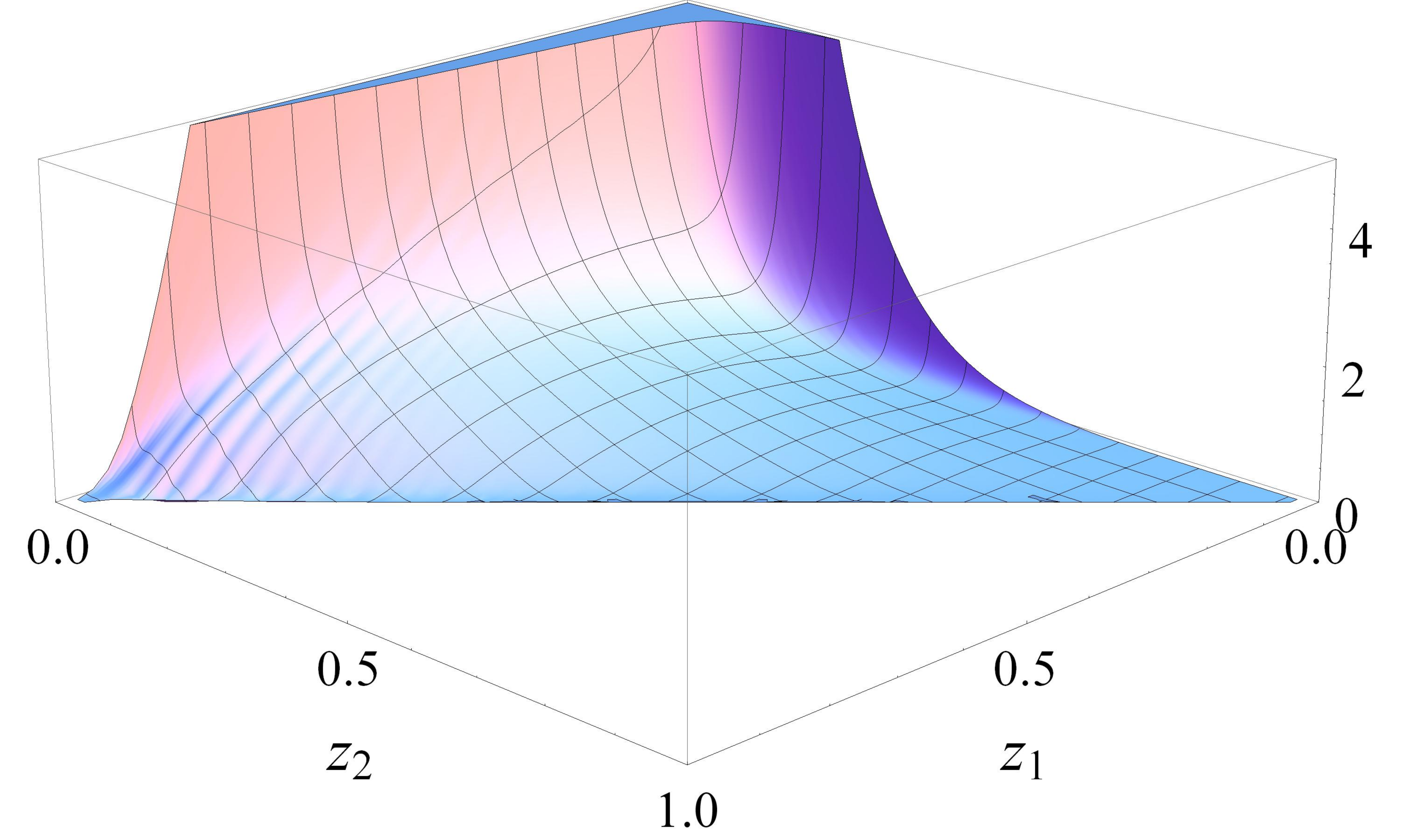}}  
\caption{$\pi^+ \pi^-$ dihadron fragmentation function for the $u$ quark at the~\subref{fig:upplpmi3dmodel} model scale~($Q_0^2=0.2 \text{ GeV}^2$) and~\subref{fig:upplpmi3devol} the evolved scale~($Q^2=4 \text{ GeV}^2$)}
\label{fig:upplpmi3d}
\end{center}
\end{figure}

Results for the SFFs and DFFs from the NJL-jet model at the model scale of $Q^2_0=0.2~\mathrm{GeV}^2$ are described in detail in Ref.~\cite{PhysRevD.85.114049}. In this paper, they are used as the input for the DFF evolution equations that will be discussed in Sections~\ref{sec:evolsff} and~\ref{sec:evoldffs}. In Fig.~\ref{fig:upplpmi3dmodel}, we present a 3-dimensional plot of $D^{\pi^+ \pi^-}_u(z_1,z_2)$, at the model scale, while in Fig.~\ref{fig:upplpmi3devol} the result for the same DFF evolved to $4~\mathrm{GeV}^2$ is shown. These plots demonstrate the effect of evolution on the DFFs, particularly where the functions achieve their peaks with respect to $z_1$ and $z_2$.

\section{Evolution of the SFFs}
\label{sec:evolsff}
To evolve the DFFs, we need to evolve the SFFs as well. This section will focus on the evolution of the SFFs, and will also serve as a simple introduction to the method used to solve the DFF evolution equations. This procedure for solving the SFF and DFF evolution equations can, of course, be used for models other than the NJL-jet model.

The single hadron fragmentation function evolution equations used in our calculations were based on those presented in Ref.~\cite{Hirai:2011si}. The evolution equations are written in the form of non-singlet quark, plus-type quark and gluon fragmentation function equations. The plus-type quark and gluon fragmentation functions are coupled and therefore need to be solved simultaneously, whereas the non-singlet quark fragmentation functions are decoupled and can be solved separately. The non-singlet~$\left(D_{q_i^-}^h(z,Q^2)\right)$ and plus-type~$\left(D_{q_i^+}^h(z,Q^2)\right)$ quark fragmentation functions are, respectively, constructed from the combinations of SFFs 
\begin{align}
\label{NSSFF}
D_{q_i^-}^h(z,Q^2) & = D_{q_i}^h(z,Q^2)-D_{\bar{q_i}}^h(z,Q^2) \nonumber\\
& =D_{q_i}^h(z,Q^2)-D_{q_i}^{\bar{h}}(z,Q^2),
\end{align}
and
\begin{align}
\label{SSFF}
D_{q_i^+}^h(z,Q^2) & = D_{q_i}^h(z,Q^2)+D_{\bar{q_i}}^h(z,Q^2)\nonumber\\
& =D_{q_i}^h(z,Q^2)+D_{q_i}^{\bar{h}}(z,Q^2),
\end{align}
where $q_i$ is the fragmenting quark. These combinations, rewritten using charge symmetry, allow for a simpler method of solving the evolution equations.

We define the variable $t$ as
\begin{equation}
\label{tvariable}
t \equiv -\frac{2}{\beta_0}\ln \left[\frac{\alpha_s(Q^2)}{\alpha_s(Q_0^2)}\right],
\end{equation}
where 
\begin{equation}
\alpha_s(Q^2)=4\pi/\left(\beta_0\ln\frac{Q^2}{\Lambda_{QCD}^2}\right),
\end{equation}
is the leading-order strong coupling constant, $\beta_0=(33-2n_f)/3$ is the one-loop $\beta$ function, $n_f$ is the number of flavors and $\Lambda_{QCD}$ is the QCD scale parameter\footnote{In this work we take $n_f=3$ and $\Lambda_{QCD}=0.25$.}. We write the evolution equations with respect to $t$ rather than $\ln Q^2$ to simplify the numerical calculation.

The QCD evolution equations for the SFFs allow us to determine the SFFs at momentum scales that vary from the scale at which they are originally defined. This is achieved by calculating the rate of change of the SFF with respect to the momentum scale. The non-singlet, plus-type and gluon leading-order~(LO) evolution equations are, respectively, given by 
\begin{multline}
\label{NSSFFevol}
 \frac{\partial}{\partial t}D_{q_i^-}^h(z,t) = \sum_j \int^1_z \frac{dy}{y}P_{q_j q_i}\left(y\right)D_{q_j^-}^h\left(\frac{z}{y},t\right),
\end{multline}
\begin{multline}
\label{SSFFevol}
\frac{\partial}{\partial t}D_{q_i^+}^h(z,t) =\int^1_z \frac{dy}{y}\\ 
\left[\sum_j P_{q_j q_i}\left(y\right)D_{q_j^+}^h\left(\frac{z}{y},t\right)+2P_{gq}(y)D_{g}^h\left(\frac{z}{y},t\right)\right],
\end{multline}
\begin{multline}
\label{GSFFevol}
\frac{\partial}{\partial t}D_{g}^h(z,t) =\int^1_z \frac{dy}{y}\\ 
\left[P_{qg}\left(y\right)\sum_j D_{q_j^+}^h\left(\frac{z}{y},t\right)+P_{gg}(y)D_{g}^h\left(\frac{z}{y},t\right)\right].
\end{multline}
The left hand sides of Eqs.~\eqref{NSSFFevol}-\eqref{GSFFevol} represent the rate of change of the corresponding SFFs with respect to $t$. The right hand sides of these equations represent the effect that a parton $j$~(either a quark of flavor $q_j$ or a gluon $g$), that emits a hadron $h$ with light-cone momentum fraction $z/y$, has on the evolution of the non-singlet~($q_i^-$), plus-type~($q_i^+$) or gluon~($g$) SFFs, through the splitting functions $P_{j i}(y)$~(obtained from Ref.~\cite{Hirai:2011si}), where $i$ is the parton for the corresponding SFF on the left hand side.

To solve Eqs.~\eqref{NSSFFevol}-\eqref{GSFFevol}, we express the derivatives as finite differences using 
\begin{equation}
\label{differential}
\frac{\partial f(t)}{\partial t} \equiv \frac{f(t_{j+1})-f(t_{j})}{\Delta t},
\end{equation}
where $f(t)$ is the corresponding SFF. We divide the range of interest for $t$ into $N_t$ steps of size $\Delta t$.

The integrals on the right hand side of the LO evolution equations are converted into sums over logarithmically disretized values of $y$~(denoted by $z_l$).
The corresponding equations for the non-singlet, plus-type and gluon fragmentation functions are, respectively, rearranged to obtain the functions at the $(k+1)\mathrm{th}$ step in $t$ such that
\begin{widetext}
\begin{align}
\label{NSSFFdisc}
D_{q_i^-}^h(z_m,t_{k+1}) & = D_{q_i^-}^h(z_m,t_{k})  +\Delta t \sum_j \sum_{l=m}^{N_z} \frac{\Delta z_l}{z_l} P_{q_j q_i}\left(z_l\right)D_{q_j^-}^h\left(\frac{z_m}{z_l},t_k\right),\\
\nonumber\\
\label{SSFFdisc}
D_{q_i^+}^h(z_m,t_{k+1}) &  = D_{q_i^+}^h(z_m,t_{k})  + \Delta t \sum_{l=m}^{N_z} \frac{\Delta z_l}{z_l} \left[\sum_j P_{q_j q_i}\left(z_l\right)D_{q_j^+}^h\left(\frac{z_m}{z_l},t_k\right)+2P_{gq}(z_l)D_{g}^h\left(\frac{z_m}{z_l},t_k\right)\right],\\
\nonumber\\
\label{GSFFdisc}
D_{g}^h(z_m,t_{k+1}) & = D_{g}^h(z_m,t_{k})  + \Delta t \sum_{l=m}^{N_z} \frac{\Delta z_l}{z_l} \left[P_{qg}\left(z_l\right)\sum_j D_{q_j^+}^h\left(\frac{z_m}{z_l},t_k\right)+P_{gg}(z_l)D_{g}^h\left(\frac{z_m}{z_l},t_k\right)\right].
\end{align}
\end{widetext}
The first term on the right sides of Eqs.~\eqref{NSSFFdisc}-\eqref{GSFFdisc} are the fragmentation functions at the $(k)\mathrm{th}$ step in $t$. The second term on the right hand side of each equation is the change in the fragmentation function from the $(k)\mathrm{th}$ step to the $(k+1)\mathrm{th}$ step in $t$. The SFF at $Q_0^2$ are inserted as the input at $k=1$, with the evolution to the next step, $t_2=t_1 +\Delta t$, calculated using the previous result. This process is repeated to obtain the SFF evolved to the chosen $Q^2$ at $t_{N_t+1}$. 

\section{Evolution of the DFFs}
\label{sec:evoldffs}
The DFF evolution equations are derived from factorization of the cross-section for the production of two hadrons in $e^+e^-$ annihilation in the $\overline{\text{MS}}$ factorization scheme in Ref.~\cite{deFlorian:2003cg}. Using jet-calculus, Ref.~\cite{Ceccopieri:2007ip} deduces the evolution equations for DFFs with an explicit dependence on the invariant mass of the hadron pairs, $M_h$, which are addressed as extended dihadron fragmentation functions. The latter are important as they relate to experimental results that include the dependence on invariant mass spectra. We concentrate on the DFF that have been integrated over the invariant mass. The LO evolution equation for DFFs, from Ref.~\cite{Ceccopieri:2007ip}, reads
\begin{widetext}
\begin{multline}
\label{DFFEVOeq}
\frac{d}{d\ln Q^2} D^{h_1 h_2}_i(z_1,z_2,Q^2) =  \frac{\alpha_s(Q^2)}{2\pi}\int^1_{z_1+z_2} \frac{du}{u^2}\,\,\, D^{h_1 h_2}_j\left(\frac{z_1}{u},\frac{z_2}{u},Q^2\right)P_{ji}(u)\\
 +  \frac{\alpha_s(Q^2)}{2\pi}   \int^{1-z_2}_{z_1} \frac{du}{u(1-u)}\,\,\,D^{h_1}_j\left(\frac{z_1}{u},Q^2\right)D^{h_2}_k\left(\frac{z_2}{1-u},Q^2\right)\hat{P}^i_{kj}(u),
\end{multline}
\end{widetext}
where $Q^2$ is the momentum scale, $\alpha_s(Q^2)$ is the strong coupling constant at the corresponding momentum scale and a sum over the repeated indices is implied. The rate at which the DFFs change with respect to $\ln Q^2$ is represented on the left hand side of Eq.~\eqref{DFFEVOeq}. The first term on the right hand side of the LO DFF evolution equation represents the effect that a parton $j$ fragmenting into two hadrons, $h_1$ and $h_2$, has on the fragmentation of parton $i$ into the two hadrons, through the splitting function $P_{j i}(u)$. The second term represents the effect of parton $i$ splitting into two partons, $j$ and $k$, that fragment separately to produce $h_1$ and $h_2$ with light-cone momentum fractions $u$ and $1-u$, respectively, through the splitting function $\hat{P}^i_{kj}(u)$. For the QCD evolution of the DFFs, both $P_{j i}(u)$ and $\hat{P}^i_{kj}(u)$ were obtained from Ref.~\cite{Ceccopieri:2007ip}.

In Eq.~\eqref{DFFEVOeq}, the parton $i$ can be either a quark, antiquark or gluon. We choose to express the evolution equations for the quark and gluon DFFs, respectively, written in terms of $t$~(Eq.~\eqref{tvariable}) as
\begin{widetext}
\begin{align}
\label{QDFF}
 \frac{d}{dt} D^{h_1 h_2}_{q_i}(z_1,z_2,t) & =  \int^1_{z_1+z_2} \frac{du}{u^2}  D^{h_1 h_2}_{q_j}\left(\frac{z_1}{u},\frac{z_2}{u},t\right)  P_{q_j q_i}(u)  +   \int^{1-z_2}_{z_1} \frac{du}{u(1-u)} D^{h_1}_g\left(\frac{z_1}{u},t\right)D^{h_2}_{q_k}\left(\frac{z_2}{1-u},t\right)\hat{P}^{q_i}_{q_k g}(u) \nonumber \\
& + \int^1_{z_1+z_2} \frac{du}{u^2} D^{h_1 h_2}_{g}\left(\frac{z_1}{u},\frac{z_2}{u},t\right)P_{g q_i}(u) + \int^{1-z_2}_{z_1} \frac{du}{u(1-u)}D^{h_1}_{q_j}\left(\frac{z_1}{u},t\right)D^{h_2}_{g}\left(\frac{z_2}{1-u},t\right)\hat{P}^{q_i}_{g q_j}(u),\\
\label{GDFF}
\frac{d}{dt} D^{h_1 h_2}_{g}(z_1,z_2,t) &  =  \int^1_{z_1+z_2} \frac{du}{u^2} D^{h_1 h_2}_{q_j}\left(\frac{z_1}{u},\frac{z_2}{u},t\right)P_{q_j g}(u)+  \int^{1-z_2}_{z_1} \frac{du}{u(1-u)}D^{h_1}_{q_j}\left(\frac{z_1}{u},t\right)D^{h_2}_{\bar{q}_j}\left(\frac{z_2}{1-u},t\right)\hat{P}^{g}_{\bar{q}_j q_j}(u)\nonumber \\
&  + \int^1_{z_1+z_2} \frac{du}{u^2} D^{h_1 h_2}_{g}\left(\frac{z_1}{u},\frac{z_2}{u},t\right)P_{g g}(u) + \int^{1-z_2}_{z_1} \frac{du}{u(1-u)}D^{h_1}_{g}\left(\frac{z_1}{u},t\right)D^{h_2}_{g}\left(\frac{z_2}{1-u},t\right)\hat{P}^{g}_{g g}(u),
\end{align}
\end{widetext}

To obtain non-singlet~$\left(D^{h_1 h_2}_{q_i^-}(z_1,z_2,t)\right)$ and plus-type~$\left(D^{h_1 h_2}_{q_i^+}(z_1,z_2,t)\right)$ quark DFFs we use the combinations
\begin{align}
\label{NSDFFC}
\hspace{-0.16cm}D^{h_1 h_2}_{q_i^-}(z_1,z_2,t)  & =  D^{h_1 h_2}_{q_i}(z_1,z_2,t) - D^{h_1 h_2}_{\bar{q}_i}(z_1,z_2,t)\nonumber \\
& = D^{h_1 h_2}_{q_i}(z_1,z_2,t) - D^{\bar{h}_1 \bar{h}_2}_{q_i}(z_1,z_2,t),
\end{align}
and

\begin{align}
\label{SDFFC}
\hspace{-0.16cm}D^{h_1 h_2}_{q_i^+}(z_1,z_2,t) & = D^{h_1 h_2}_{q_i}(z_1,z_2,t) + D^{h_1 h_2}_{\bar{q}_i}(z_1,z_2,t) \nonumber\\
& = D^{h_1 h_2}_{q_i}(z_1,z_2,t) + D^{\bar{h}_1 \bar{h}_2}_{q_i}(z_1,z_2,t),
\end{align}
respectively. The combination of terms on the second line of each equation has been rewritten using charge symmetry and this is the form that is employed to solve the LO DFF evolution equations.

Using Eqs.~\eqref{NSDFFC} and~\eqref{SDFFC}, we write the evolution equations in terms of the non-singlet quark, plus-type quark and gluon DFFs as 
\begin{widetext}
\begin{align}
\label{NSDFFv2} 
\frac{d}{dt} D^{h_1 h_2}_{q^-_i}(z_1,z_2,t)  = & \sum_{j=u,d,s} \int^1_{z_1+z_2} \frac{du}{u^2}\,\,\, D^{h_1 h_2}_{q_j^-}\left(\frac{z_1}{u},\frac{z_2}{u},t\right)P_{q_j q_i}(u)\nonumber \\
&+ \sum_{k=u,d,s}\int^{1-z_2}_{z_1} \frac{du}{u(1-u)}\,\,\,D^{h_1}_g\left(\frac{z_1}{u},t\right)D^{h_2}_{q_k^-}\left(\frac{z_2}{1-u},t\right)\hat{P}^{q_i}_{q_k g}(u)\nonumber \\
& +\sum_{j=u,d,s}\int^{1-z_2}_{z_1} \frac{du}{u(1-u)}\,\,\,D^{h_1}_{q_j^-}\left(\frac{z_1}{u},t\right)D^{h_2}_{g}\left(\frac{z_2}{1-u},t\right)\hat{P}^{q_i}_{g q_j}(u),\\
\nonumber\\
\label{SDFFv2}
\frac{d}{dt} D^{h_1 h_2}_{q^+_i}(z_1,z_2,t) =& \sum_{j=u,d,s} \int^1_{z_1+z_2} \frac{du}{u^2}\,\,\, D^{h_1 h_2}_{q_j^+}\left(\frac{z_1}{u},\frac{z_2}{u},t\right)P_{q_j q_i}(u)+2\int^1_{z_1+z_2} \frac{du}{u^2}\,\,\, D^{h_1 h_2}_{g}\left(\frac{z_1}{u},\frac{z_2}{u},t\right)P_{g q_i}(u)\nonumber \\
&  + \sum_{k=u,d,s}\int^{1-z_2}_{z_1} \frac{du}{u(1-u)}\,\,\,D^{h_1}_g\left(\frac{z_1}{u},t\right)D^{h_2}_{q_k^+}\left(\frac{z_2}{1-u},t\right)\hat{P}^{q_i}_{q_k g}(u)\nonumber \\
&  +\sum_{j=u,d,s}\int^{1-z_2}_{z_1} \frac{du}{u(1-u)}\,\,\,D^{h_1}_{q_j^+}\left(\frac{z_1}{u},t\right)D^{h_2}_{g}\left(\frac{z_2}{1-u},t\right)\hat{P}^{q_i}_{g q_j}(u),
\end{align}
\begin{align}
\label{GDFFv2}
\frac{d}{dt} D^{h_1 h_2}_{g}(z_1,z_2,t)  =  & \sum_{j=u,d,s,\bar{u},\bar{d},\bar{s}}\int^1_{z_1+z_2} \frac{du}{u^2}\,\,\, D^{h_1 h_2}_{q_j}\left(\frac{z_1}{u},\frac{z_2}{u},t\right)P_{q_j g}(u)+ \int^1_{z_1+z_2} \frac{du}{u^2}\,\,\, D^{h_1 h_2}_{g}\left(\frac{z_1}{u},\frac{z_2}{u},t\right)P_{g g}(u)\nonumber \\
& +  \sum_{j=u,d,s,\bar{u},\bar{d},\bar{s}}\int^{1-z_2}_{z_1} \frac{du}{u(1-u)}\,\,\,D^{h_1}_{q_j}\left(\frac{z_1}{u},t\right)D^{h_2}_{\bar{q}_j}\left(\frac{z_2}{1-u},t\right)\hat{P}^{g}_{\bar{q}_j q_j}(u)\nonumber \\
& + \int^{1-z_2}_{z_1} \frac{du}{u(1-u)}\,\,\,D^{h_1}_{g}\left(\frac{z_1}{u},t\right)D^{h_2}_{g}\left(\frac{z_2}{1-u},t\right)\hat{P}^{g}_{g g}(u),\nonumber \\
\nonumber\\
 = & \sum_{j=u,d,s}\int^1_{z_1+z_2} \frac{du}{u^2}\,\,\,\left[D^{h_1 h_2}_{q_j^+}\left(\frac{z_1}{u},\frac{z_2}{u},t\right)P_{q_j g}(u)\right]+ \int^1_{z_1+z_2} \frac{du}{u^2}\,\,\, D^{h_1 h_2}_{g}\left(\frac{z_1}{u},\frac{z_2}{u},t\right)P_{g g}(u)\nonumber \\
& +  \sum_{j=u,d,s,\bar{u},\bar{d},\bar{s}} \int^{1-z_2}_{z_1} \frac{du}{u(1-u)}\,\,\,D^{h_1}_{q_j}\left(\frac{z_1}{u},t\right)D^{h_2}_{\bar{q}_j}\left(\frac{z_2}{1-u},t\right)\hat{P}^{g}_{\bar{q}_j q_j}(u)\nonumber \\
& +  \int^{1-z_2}_{z_1} \frac{du}{u(1-u)}\,\,\,D^{h_1}_{g}\left(\frac{z_1}{u},t\right)D^{h_2}_{g}\left(\frac{z_2}{1-u},t\right)\hat{P}^{g}_{g g}(u).
\end{align}
\end{widetext}
For clarity, we show the sums over the repeated indices and use Eq.~\eqref{GDFFv2} to display how the combinations are applied to simplify the equations. The non-singlet quark evolution equation is decoupled from the plus-type quark and gluon DFFs and can be evolved separately from them. Using Eq.~\eqref{differential} and converting integrals into sums over logarithmically discretized values of $u$, expressions for the DFFs evolved to the $(k+1)\mathrm{th}$ step in $t$ can be obtained, producing results analogous to Eqs.~\eqref{NSSFFdisc}-\eqref{GSFFdisc}.

\section{Results}
\label{sec:evolresults}
In this section we present the results comparing the model scale DFFs with those evolved to $Q^2=4~\mathrm{GeV}^2$ for $u\to \pi^+ \pi^-$, $u\to \pi^+ K^-$ and $q\to K^+ K^-$, where $q=u,d,s$. The first subsection explores the evolution of $D^{\pi^+ \pi^-}_u$ by comparing the model and evolved DFFs at particular values of $z_1$ or $z_2$, while the second subsection focuses on favored and unfavored hadron emission in the evolution of $D^{\pi^+ K^-}_u$. Finally, the last subsection demonstrates the evolution of $D^{K^+ K^-}_q$ for $q=u$, $d$ or $s$. 

\subsection{$Q^2$ evolution of $D^{\pi^+ \pi^-}_u$}
\label{sec:upplpmievol}
\begin{figure}[t]

\begin{center}
\subfigure[$\,z_1=0.5$: $z_2\,D^{\pi^+ \pi^-}_u$]{\label{fig:upplpmiz105}
\includegraphics[width=0.48\textwidth]{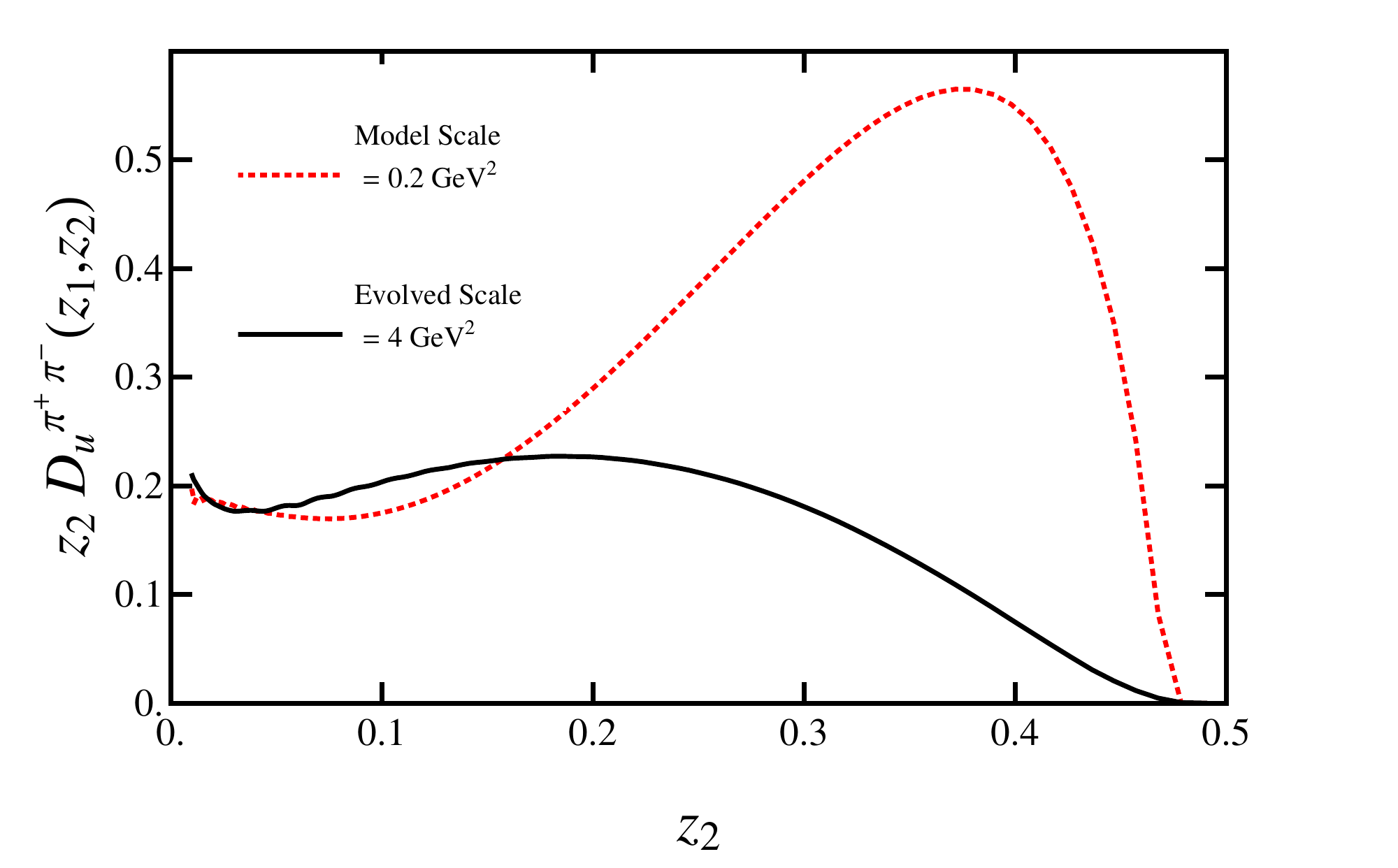}}
\subfigure[$\,z_2=0.5$: $z_1\,D^{\pi^+ \pi^-}_u$]{\label{fig:upplpmiz205}
\includegraphics[width=0.48\textwidth]{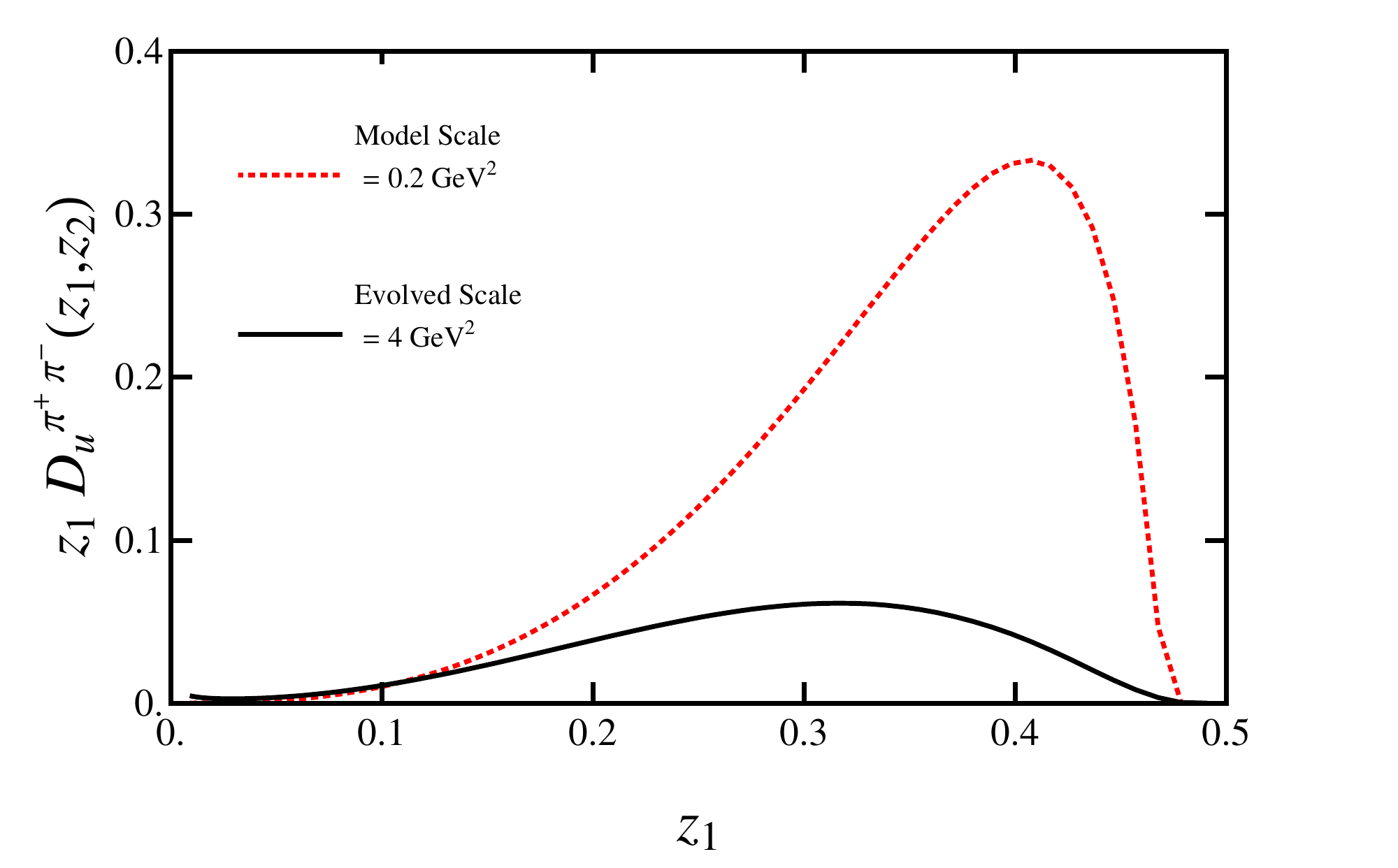}}  
\caption{$\pi^+ \pi^-$ dihadron fragmentation functions for the $u$ quark at the model scale~($Q_0^2=0.2 \text{ GeV}^2$, shown by dotted red line) and the evolved scale~($Q^2=4 \text{ GeV}^2$, shown by solid black line) for \subref{fig:upplpmiz105} $z_1=0.5$ and \subref{fig:upplpmiz205} $z_2=0.5$.}
\label{fig:upplpmi}
\end{center}
\end{figure}

We consider the DFF for an up quark fragmenting to $\pi^+$ and $\pi^-$. When the up quark fragments to $\pi^+$, for which it is the favored emission channel, it produces a down quark, which has the favored emission channel to $\pi^-$. Since both emissions are favored channels for the detected hadrons in this quark cascade, the DFF has sizeable peaks in the higher $z_2$ and $z_1$ regions for $z_1=0.5$~(Fig.~\ref{fig:upplpmiz105}) and $z_2=0.5$~(Fig.~\ref{fig:upplpmiz205}), respectively. For $D^{\pi^+ \pi^-}_u$, the second term of Eq.~\eqref{DFFcv} is zero~(because $\hat{d}^{\pi^-}_u=0$) and the integral term is small, so this DFF is dominated by the first term of Eq.~\eqref{DFFcv}. The model scale plot for $D^{\pi^+ \pi^-}_u$ fixed at $z_1=0.5$~(Fig.~\ref{fig:upplpmiz105}) has the shape of a favored single hadron fragmentation function since fixing $z_1$ effectively makes the first term on the right hand side of Eq.~\eqref{DFFcv} a constant multiplied by the favored fragmentation $D^{\pi^-}_d$. For $z_2$ fixed at $0.5$~(Fig~\ref{fig:upplpmiz205}), the model scale $D^{\pi^+ \pi^-}_u$ is shaped by the elementary quark fragmentation function $\hat{d}^{\pi^+}_u$, resulting in a peak at higher $z_1$, while having a very small contribution at low values of $z_1$. After evolution of the DFF, there is a reduction in magnitude and a shift in the peak towards the low $z_2$ region for $z_1=0.5$~(Fig.~\ref{fig:upplpmiz105}). When $z_2$ is fixed at $0.5$~(Fig.~\ref{fig:upplpmiz205}), the magnitude of the DFF is reduced and the peak value shifts towards the low $z_1$ region. Both plots in Figs.~\ref{fig:upplpmi} display a range of values at low $z$ where the evolved DFF obtains a larger magnitude than the model scale DFF. At higher momentum scales, the low $z_1$ and $z_2$ regions of the DFFs grow in magnitude because they can access the gluon emission channel.

\begin{figure}[t]
\begin{center}
\subfigure[$\,z_1=0.2$: $z_2\,D^{\pi^+ \pi^-}_u$]{\label{fig:upplpmiz102}
\includegraphics[width=0.48\textwidth]{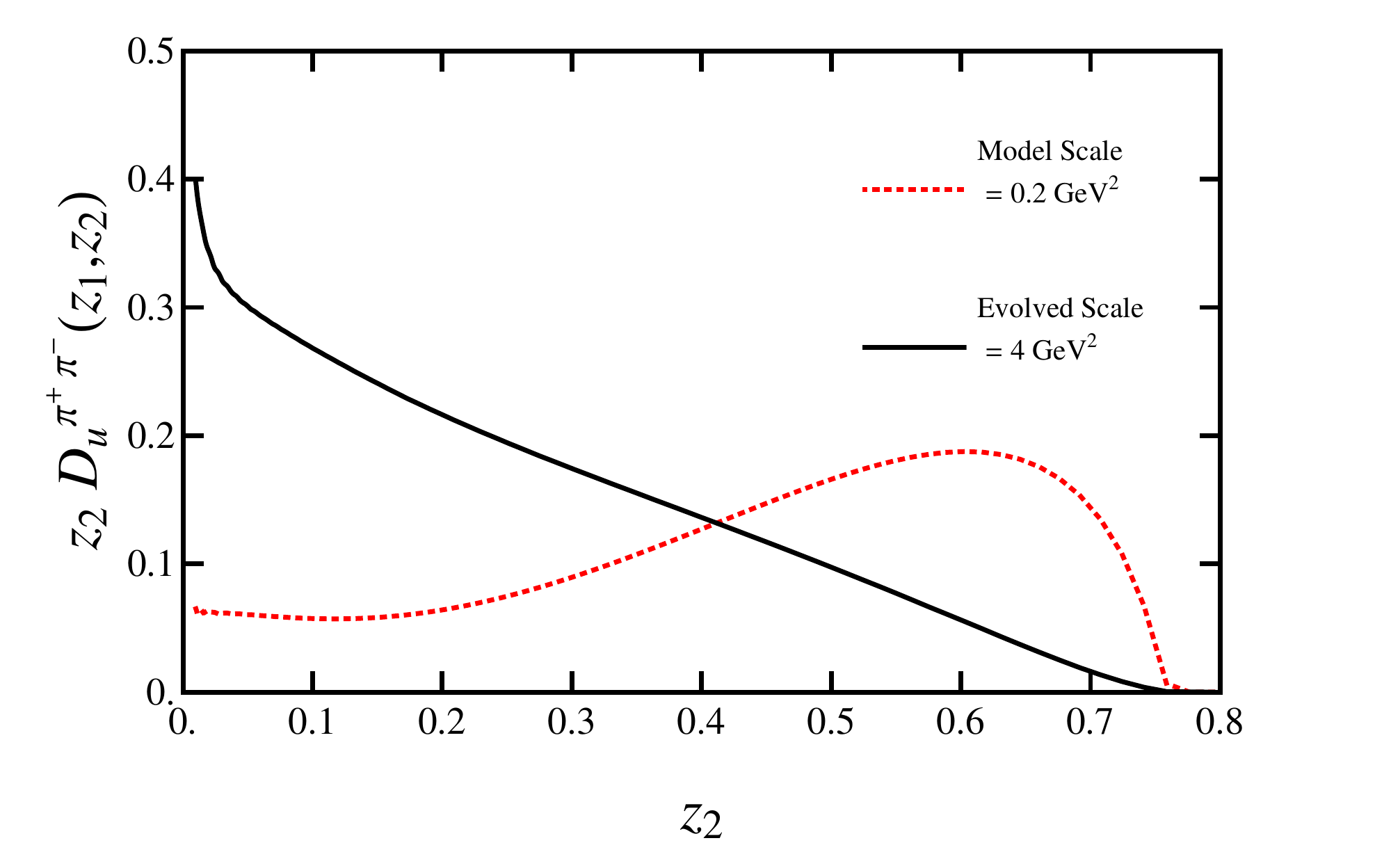}}
\subfigure[$\,z_2=0.2$: $z_1\,D^{\pi^+ \pi^-}_u$]{\label{fig:upplpmiz202}
\includegraphics[width=0.48\textwidth]{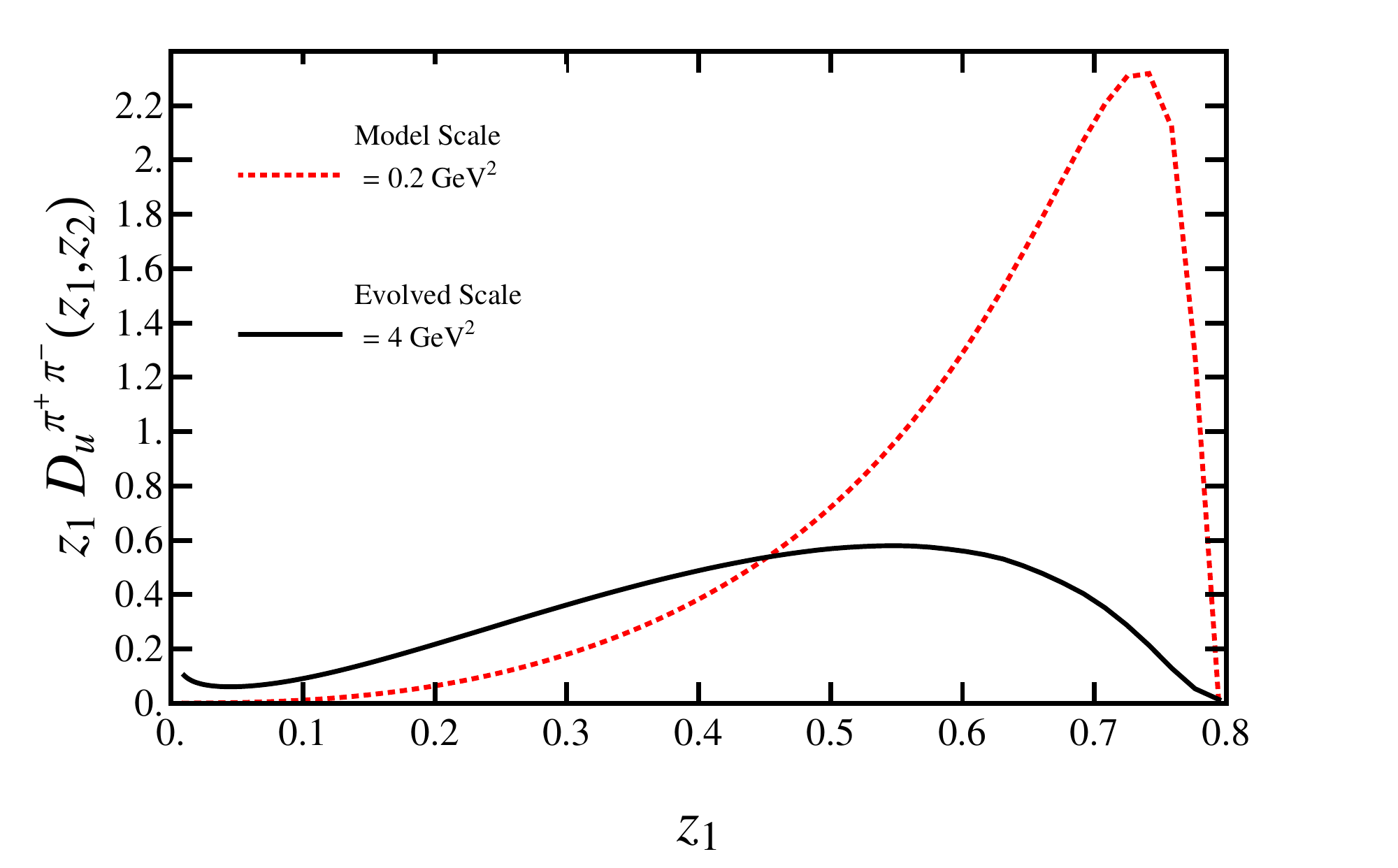}}
\caption{$\pi^+ \pi^-$ dihadron fragmentation functions for the $u$ quark at the model scale~($Q_0^2=0.2 \text{ GeV}^2$, shown by dotted red line) and the evolved scale~($Q^2=4 \text{ GeV}^2$, shown by solid black line) for \subref{fig:upplpmiz105} $z_1=0.2$ and \subref{fig:upplpmiz205} $z_2=0.2$.}
\label{fig:upplpmi02}
\end{center}
\end{figure}

We present the results for $z_1$ and $z_2$ fixed to $0.2$ in Figs.~\ref{fig:upplpmiz102} and \ref{fig:upplpmiz202}, respectively, to investigate the DFF at low fixed light-cone momentum fraction. The structure of the model scale $D^{\pi^+ \pi^-}_u$ for $z_1=0.2$, shown in Fig.~\ref{fig:upplpmiz102}, is similar in shape to that of the model scale DFF at $z_1=0.5$, having the peak in the higher $z_2$, except it is spread out more and has a lower peak value. At $z_2=0.2$, the structure of the model scale DFF is again similar to the corresponding $z_2=0.5$ plot in Fig.~\ref{fig:upplpmiz205}, being small in the low $z_1$ region and having a large peak in the higher $z_1$ region, which is rather narrow. Evolution of the DFF results in a shift of the peak towards the lower $z$ regions, with the magnitude of the evolved $D^{\pi^+ \pi^-}_u$ becoming larger than the model scale $D^{\pi^+ \pi^-}_u$ at mid-range values of the allowed light-cone momentum fraction; rather than in the lower range of values that was observed for the $z_1$ and $z_2$ fixed to $0.5$ results. The shape of the evolved $D^{\pi^+ \pi^-}_u$ for $z_2=0.2$ in Fig.~\ref{fig:upplpmiz202} appears very similar to that of the evolved $D^{\pi^+ \pi^-}_u$ for $z_2=0.5$ in Fig.~\ref{fig:upplpmiz205}, whereas the shape for the evolved $D^{\pi^+ \pi^-}_u$ for $z_1=0.2$~(Fig.~\ref{fig:upplpmiz202}) is quite different to the corresponding result at $z_1=0.5$ in Fig.~\ref{fig:upplpmiz105}. Instead of the concave structure at $z_1=0.5$ shown in Fig.~\ref{fig:upplpmiz105}, at $z_1=0.2$ (Fig.~\ref{fig:upplpmiz102}) the evolved DFF has a large contribution at low $z_2$ and steadily decreases as $z_2$ increases.

\subsection{$Q^2$ evolution of $D^{\pi^+ K^-}_u$}
\label{sec:upplkmievol}
\begin{figure}[t]

\begin{center}

\subfigure[$\,z_1=0.5$: $z_2\,D^{\pi^+ K^-}_u$]{\label{fig:upplkmiz105}
\includegraphics[width=0.48\textwidth]{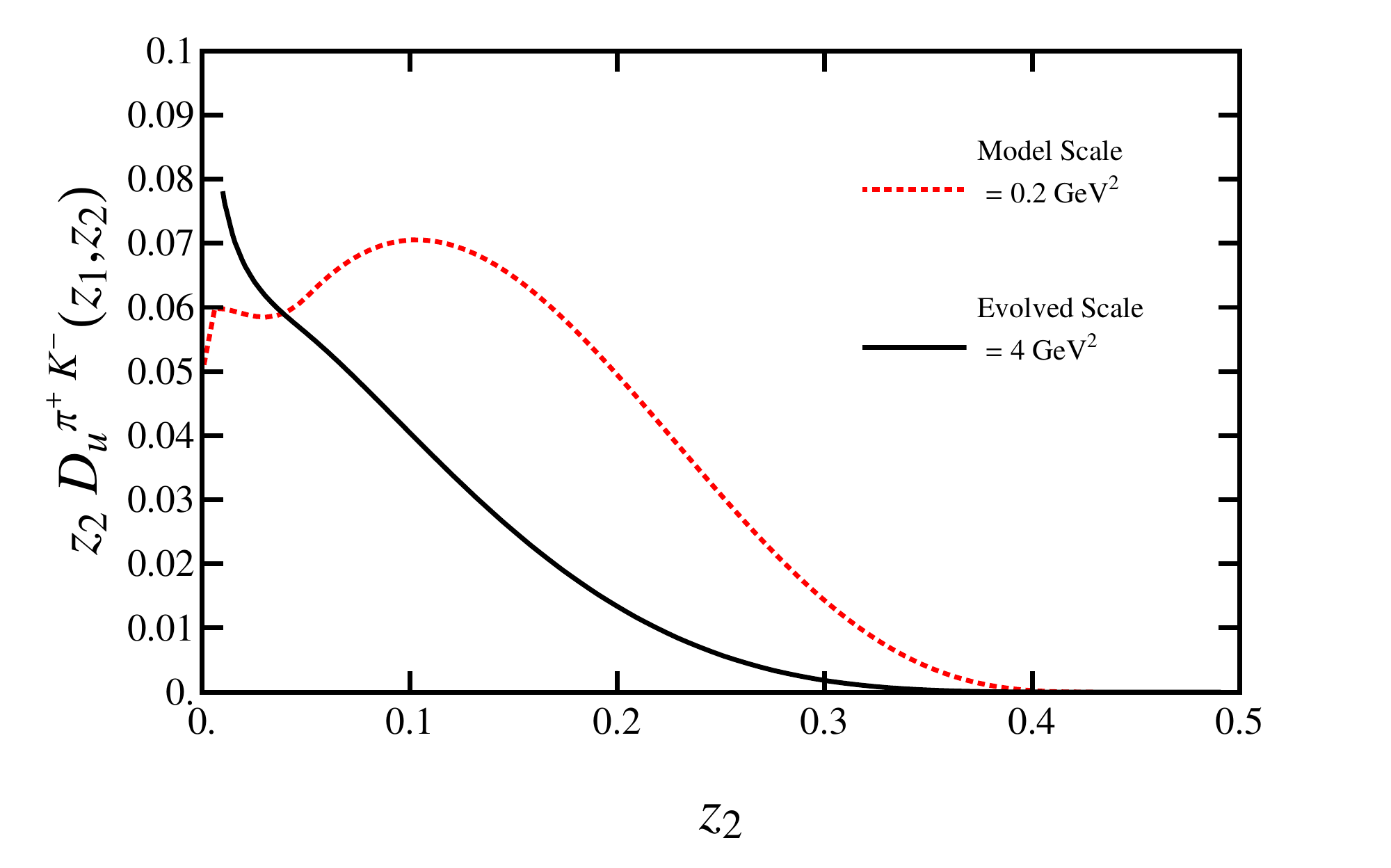}}
\subfigure[$\,z_2=0.5$: $z_1\,D^{\pi^+ K^-}_u$]{\label{fig:upplkmiz205}
\includegraphics[width=0.48\textwidth]{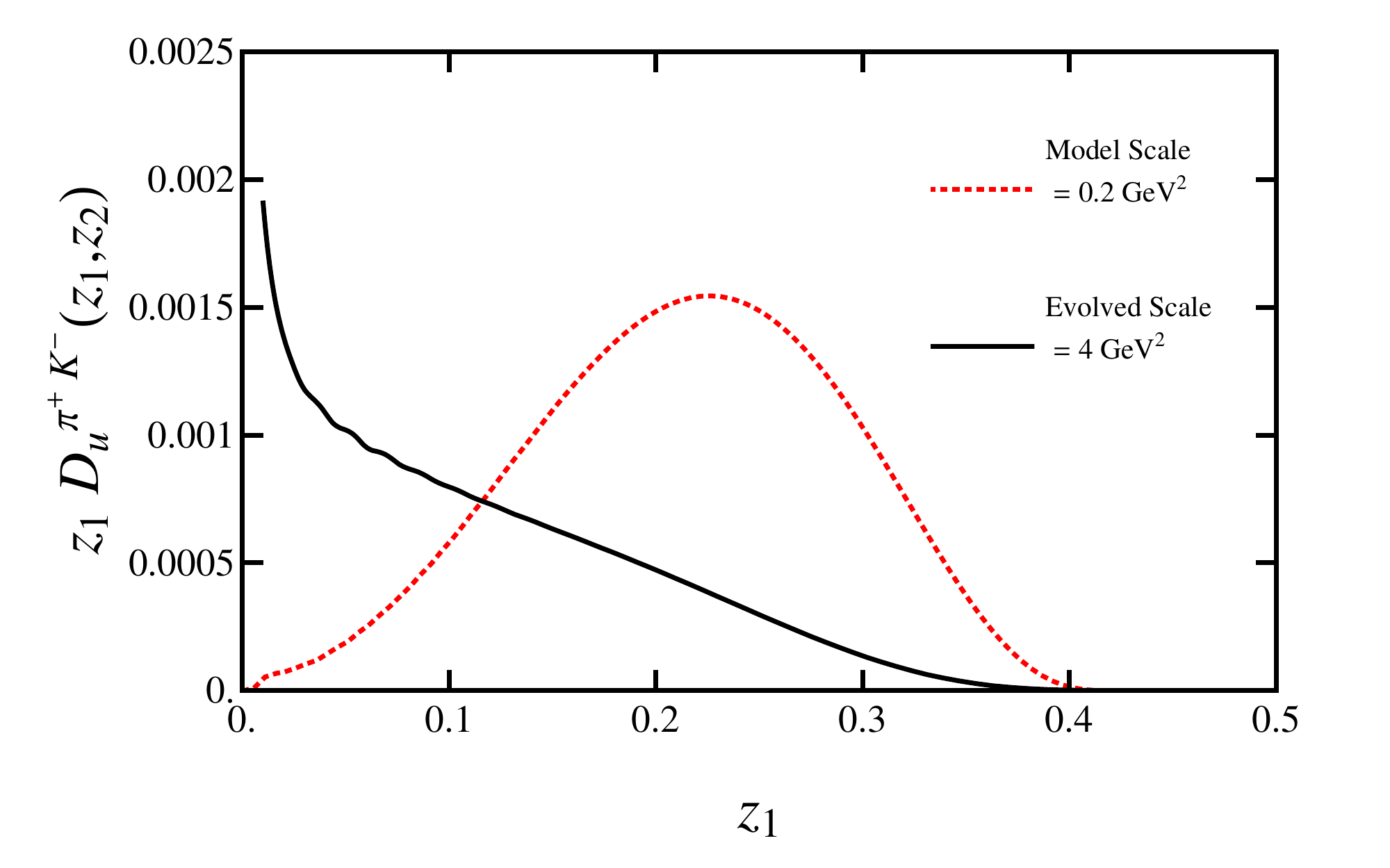}}  
\caption{$\pi^+ K^-$ dihadron fragmentation functions for the $u$ quark at the model scale~($Q_0^2=0.2 \text{ GeV}^2$, shown by dotted red line) and the evolved scale~($Q^2=4 \text{ GeV}^2$, shown by solid black line) for \subref{fig:upplkmiz105} $z_1=0.5$ and \subref{fig:upplkmiz205} $z_2=0.5$.}
\label{fig:upplkmi}
\end{center}
\end{figure}

In Figs.~\ref{fig:upplkmi} we present the results for $D^{\pi^+ K^-}_u$, where the up quark is a favored channel for $\pi^+$ emission, but the remnant down quark is an unfavored channel for $K^-$ emission. At the model scale, $D^{\pi^+ K^-}_u$ shows no contribution in the large $z_2$ and $z_1$ regions for $z_1$~(Fig.~\ref{fig:upplkmiz105}) and $z_2$~(Fig.~\ref{fig:upplkmiz205}) fixed at $0.5$, respectively. For $D^{\pi^+ K^-}_u$ at the model scale the second term of Eq.~\eqref{DFFcv} is zero~(because $\hat{d}^{K^-}_u=0$) and the integral term is small, so $D^{\pi^+ K^-}_u$ is dominated by the first term of Eq.~\eqref{DFFcv}. 
In Fig.~\ref{fig:upplkmiz105}, the model scale DFF has the structure of the unfavored $D^{K^-}_d$, while also being suppressed in magnitude by $\hat{d}^{\pi^+}_u(z_1=0.5)$, which achieves its peak value in the high $z_1$ region while vanishing in the low $z_1$ region. For $z_2=0.5$~(Fig~\ref{fig:upplkmiz205}), the model scale DFF shows a very small magnitude for values of $z_1$ because of the combination of $\hat{d}^{\pi^+}_u$ multiplied by $D^{K^-}_d$. Elementary fragmentation functions for favored emission channels are very small in the low $z$ region, and achieve large peak values in the high $z$ region. This forces $D^{\pi^+ K^-}_u(z_1,z_2)$ to have a very small magnitude in the low $z_1$ region as it is dependent on $\hat{d}^{\pi^+}_u(z_1)$. $D^{K^-}_{d}$ is an unfavored SFF and therefore is constructed by the integral term on the right hand side of Eq.~\eqref{EQ_SINGEL_FRAG} because the first term equals zero. Unfavored SFFs peak in the low $z$ region and have very small magnitudes in the medium to high $z$ region. Both of these effects combine to cause the resultant low peak in the middle of the allowed region of $z_1$. 

When the DFF is evolved there is a shift towards the low $z_2$ region for $z_1$ fixed at $0.5$~(Fig.~\ref{fig:upplkmiz105}) and towards the low $z_1$ region for $z_2$ fixed at $0.5$~(Fig.~\ref{fig:upplkmiz205}). We observe that the evolved DFF in Figs.~\ref{fig:upplkmi} has a larger magnitude in the low $z_1$ and $z_2$ regions, while steadily decreasing as $z_1$ and $z_2$ increase. This is quite different to the results shown in Figs.~\ref{fig:upplpmi}, where there is either a large contribution for almost all the allowed range of values of $z_2$~(Fig.~\ref{fig:upplpmiz105}) or a substantial peak still in the higher $z_1$ values with the magnitude of the DFF decreasing as $z_1$ is decreased. In both those cases, the DFF is largest away from the low values of $z_2$ and $z_1$. This effect could be caused by the down quark, which is produced in both fragmentations after the up quark fragments to $\pi^+$, being an unfavored emission channel for $K^-$, as opposed to the favored emission channel for $\pi^-$. The favored emission channel loses magnitude at higher $z_1$ as the momentum scale is increased, while the unfavored emission channels, which have no higher $z_1$ peak, increase at lower $z_1$ due to the greater access to the gluon emission channel.

\subsection{$Q^2$ evolution of $D^{K^+ K^-}_q$ for q = u, d or s}
\label{sec:qkplkmievol}
\begin{figure}[H!t]
\begin{center}
\subfigure[$\,z_1=0.5$: $z_2\,D^{K^+ K^-}_u$]{\label{fig:ukplkmiz105}
\includegraphics[width=0.48\textwidth]{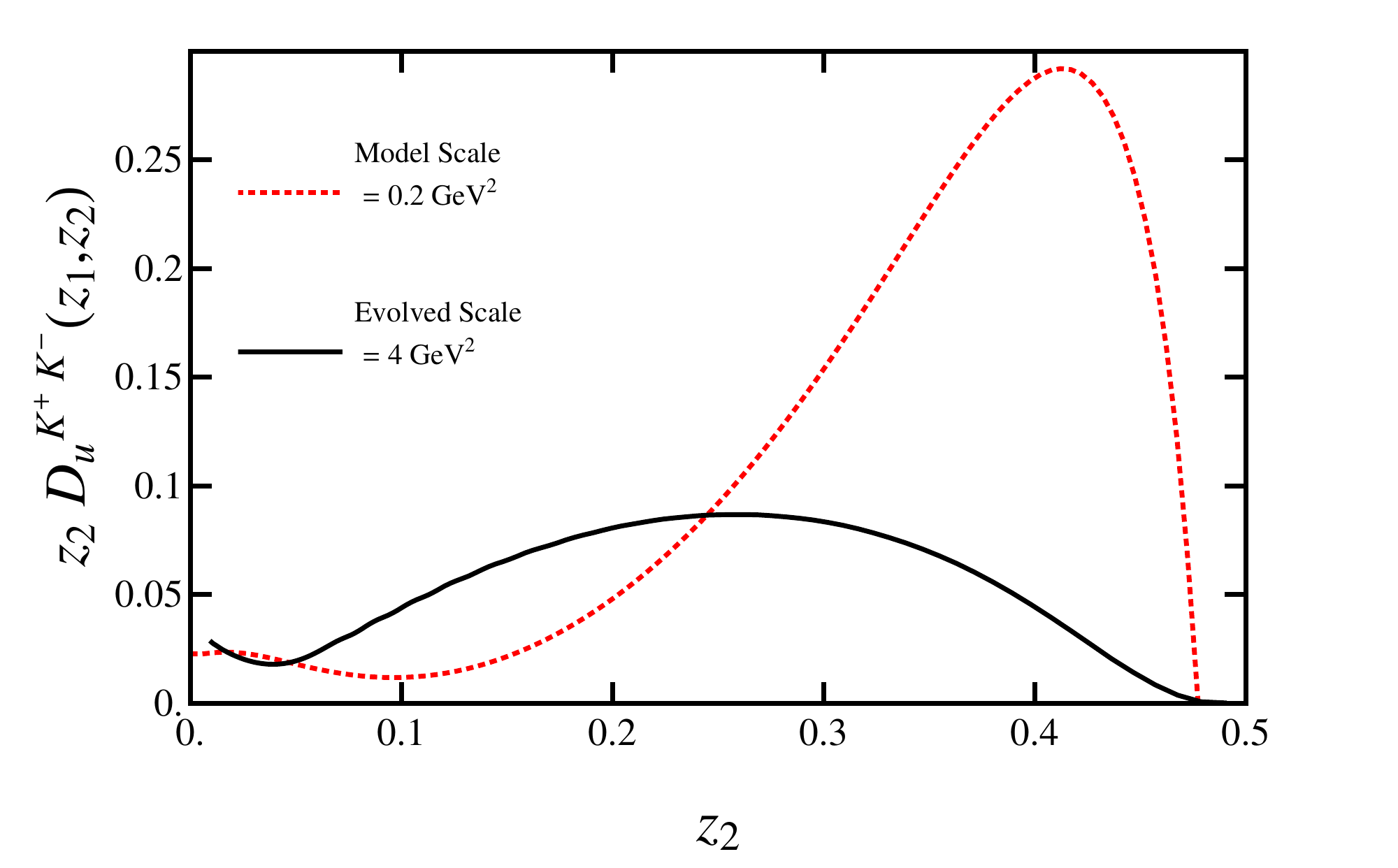}}
\subfigure[$\,z_1=0.5$: $z_2\,D^{K^+ K^-}_d$]{\label{fig:dkplkmiz105}
\includegraphics[width=0.48\textwidth]{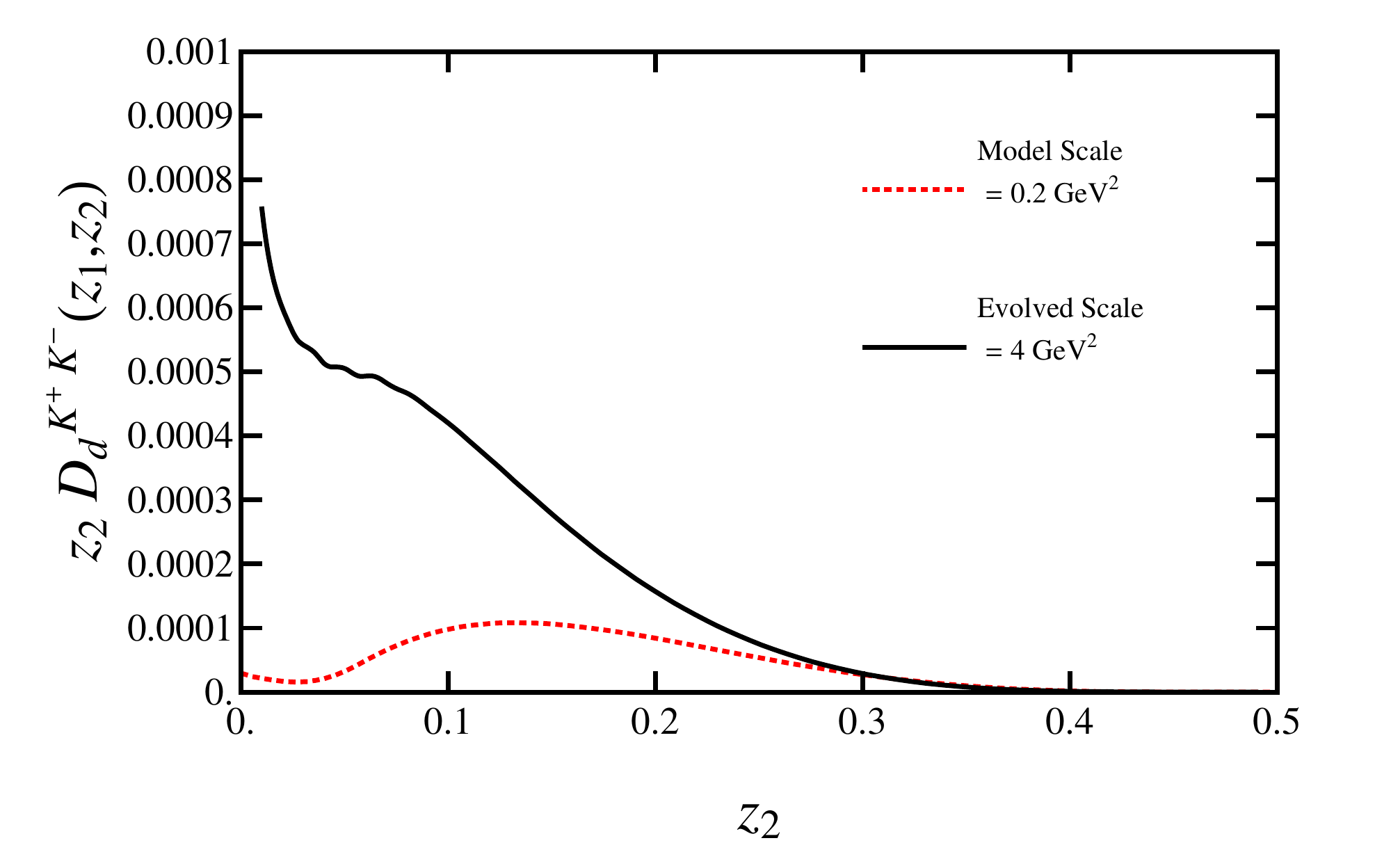}}
\subfigure[$\,z_1=0.5$: $z_2\,D^{K^+ K^-}_s$]{\label{fig:skplkmiz105}
\includegraphics[width=0.48\textwidth]{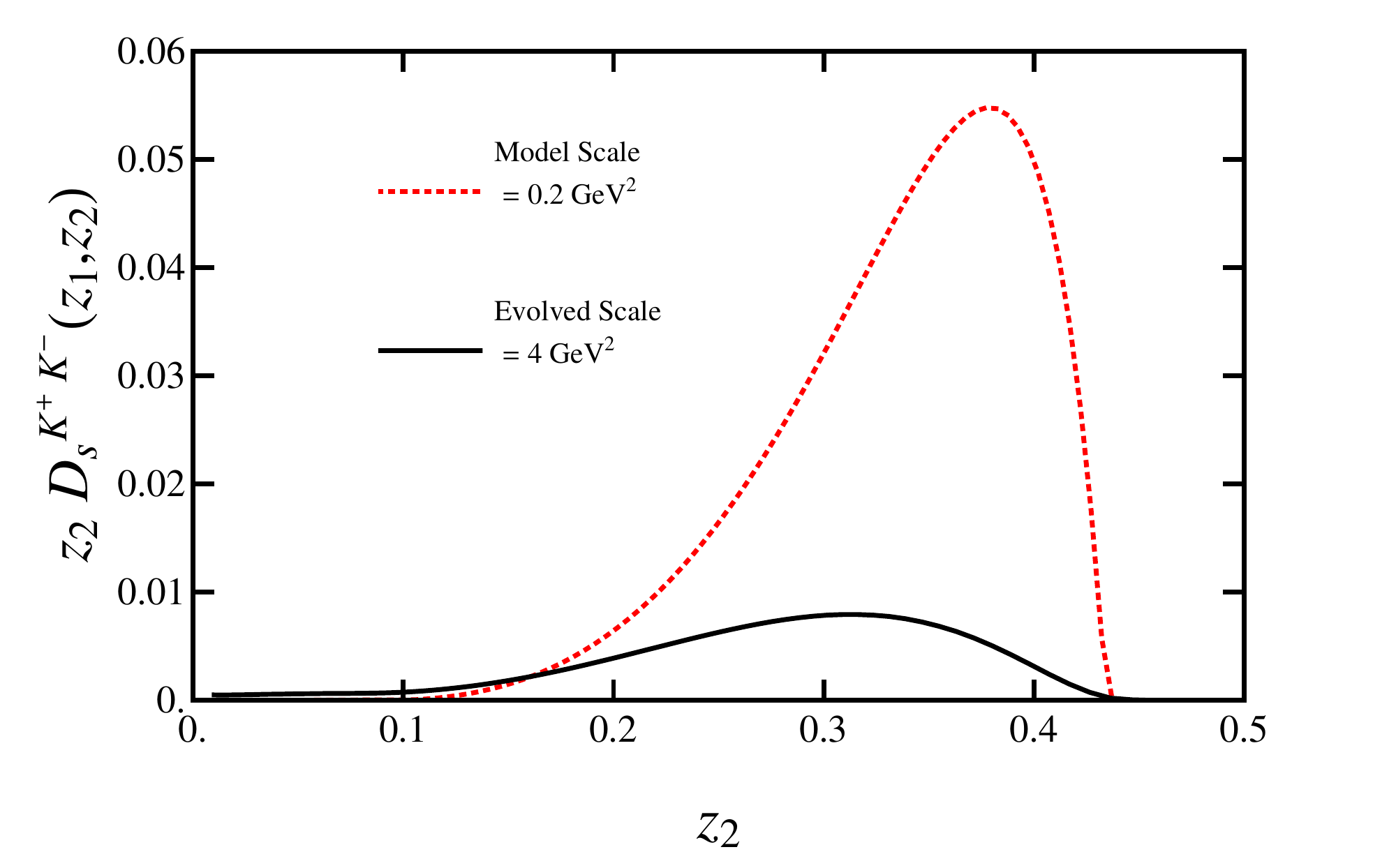}}
\caption{$K^+ K^-$ dihadron fragmentation functions for $z_1$ fixed to $0.5$ at the model scale~($Q_0^2=0.2 \text{ GeV}^2$, shown by dotted red line) and the evolved scale~($Q^2=4 \text{ GeV}^2$, shown by solid black line) for a fragmenting \subref{fig:ukplkmiz105} $u$ quark,~\subref{fig:dkplkmiz105} $d$ quark and \subref{fig:skplkmiz105} $s$ quark.}
\label{fig:kplkmi}
\end{center}
\end{figure}
We now consider $D^{K^+ K^-}_q$ for $q=u$~(Fig.~\ref{fig:ukplkmiz105}), $d$~(Fig.~\ref{fig:dkplkmiz105}) or $s$~(Fig.~\ref{fig:skplkmiz105}). The $q=u$ and $q=s$ DFFs both have large peaks in the high $z_2$ region at the model scale since both are favored fragmentation channels in the driving function of $D^{K^+ K^-}_q$. The first term on the right hand side of Eq.~\eqref{DFFcv} produces most of the magnitude of the model scale $D^{K^+ K^-}_u$ because the second term equates to zero and the intergral term is small. $D^{K^+ K^-}_s$ emerges from the second term on the right hand side of Eq.~\eqref{DFFcv} because the first term equates to zero and the integral term of the DFF is small. The first term on the right hand side of Eq.~\eqref{DFFcv} for $D^{K^+ K^-}_u$ contains the elementary quark fragmentation function for the fragmentation from an up quark to $K^+$ as a function of $z_1$, multiplied by $D^{K^-}_s(z_2/(1-z_1))/(1-z_1)$. For $z_1$ fixed to $0.5$, this term simplifies to a constant multiplied by $D^{K^-}_s(z_2/(1-z_1))$. However, for DFFs such as $D^{K^+ K^-}_s$, which emerge from the second term on the right hand side of Eq.~\eqref{DFFcv}, fixing $z_1$ to $0.5$ restricts $\hat{d}^{K^-}_s(z_2)$ to values of $z_2$ less than $0.5$. This suppresses the term considerably since the $z_2>0.5$ region of $\hat{d}^{K^-}_s(z_2)$ is where the function achieves its larger values. This is why the $u\to K^+K^-$ DFF is larger than the $s\to K^+K^-$ DFF when $z_1$ is fixed to $0.5$. 

After QCD evolution, the DFFs for fragmenting up~(Fig.~\ref{fig:ukplkmiz105}) and strange~(Fig.~\ref{fig:skplkmiz105}) quarks at $z_1=0.5$ show the shift of the peak value to the lower $z_2$ region, with $D^{K^+ K^-}_u$ having a structure similar to that seen for the evolved $D^{\pi^+ \pi^-}_u$ at $z_1=0.5$~(Fig.~\ref{fig:upplpmiz105}), while $D^{K^+ K^-}_s$ has a structure similar to that of the evolved $D^{\pi^+ \pi^-}_u$ at $z_2=0.5$~(Fig.~\ref{fig:upplpmiz205}). For $D^{K^+ K^-}_d$, the model scale plot is very small compared to $D^{K^+ K^-}_u$ and $D^{K^+ K^-}_s$, since it is unfavored for both detected hadrons. When the momentum scale is evolved up to $4~\mathrm{GeV}^2$, $D^{K^+ K^-}_d$ increases in the low $z_2$ region for $z_1$ fixed to $0.5$, because of the effects of gluon fragmentation.

\section{Comparison with other work}
\label{sec:comparison}
With very little in the way of DFFs from experiments being available for comparison, we look to compare our results with the work presented in Ref.~\cite{Majumder:2004br}. We will first show that using our code and their parameterized DFFs as initial conditions, we do indeed obtain solutions comparable with those presented in Ref.~\cite{Majumder:2004br} when evolved to $Q^2=109 \mathrm{~GeV}^2$. We also present our data evolved to a range of different scales for $D^{\pi^+ \pi^-}_u$.

First, we briefly describe the procedures used in Ref.~\cite{Majumder:2004br}. The evolution equations used there are the those of Eqs.~\eqref{QDFF} and \eqref{GDFF}, with only minor rewriting of terms in the equations. For the gluon DFF evolution equations, the difference in the equations arises from alternate definitions of the functions. In Ref.~\cite{Majumder:2004br}, the DFF is taken to be identical for the up, down and strange quarks, and so the gluon evolution equation term involving these functions is written with the function multiplied by a factor of $2n_f$, whereas the DFFs in our approach differ and so we sum over each flavor.  Similar reasoning is used for the other terms in the gluon evolution equation. To obtain the initial DFF at $Q^2=2 \mathrm{~GeV}^2$, the authors of Ref.~\cite{Majumder:2004br} simulate three million dijet events, distributed equally over the number of flavors~($n_f=3$), using JETSET. The resultant DFFs are parameterized by fitting to a functional form:
\begin{align}
D(z_1,z_2) =& N z_1^{\alpha_1} z_2^{\alpha_2}(z_1+z_2)^{\alpha_3}&\nonumber\\
&\times(1-z_1)^{\beta_1} (1-z_2)^{\beta_2}(1-z_1-z_2)^{\beta_3},
\end{align}
where $N$, $\alpha_1$, $\alpha_2$, $\alpha_3$, $\beta_1$, $\beta_2$ and $\beta_3$ are the parameters fitted by minimizing the logarithm of $\chi^2$. The fit describes the JETSET results better at larger values of $z_1$ and $z_2$, while not reproducing the results well for low values of $z_1$ and $z_2$. Values for the parameters are provided for the quark and gluon DFFs for momentum scales of $Q^2=2 \mathrm{~GeV}^2$ and $Q^2=109\mathrm{~GeV}^2$.  The SFFs used are obtained from the parameterization in Ref.~\cite{PhysRevD.52.4947}. The DFFs are QCD evolved from the initial scale of $Q^2=2 \mathrm{~GeV}^2$, and results are presented for several values of $Q^2$, including $Q^2=109\mathrm{~GeV}^2$.

\begin{figure}[H!t]
\begin{center}
\subfigure[$\,z_1=0.5$: $z_2\,D^{\pi^+ \pi^-}_u$]{\label{fig:compareupppm}
\includegraphics[width=0.48\textwidth]{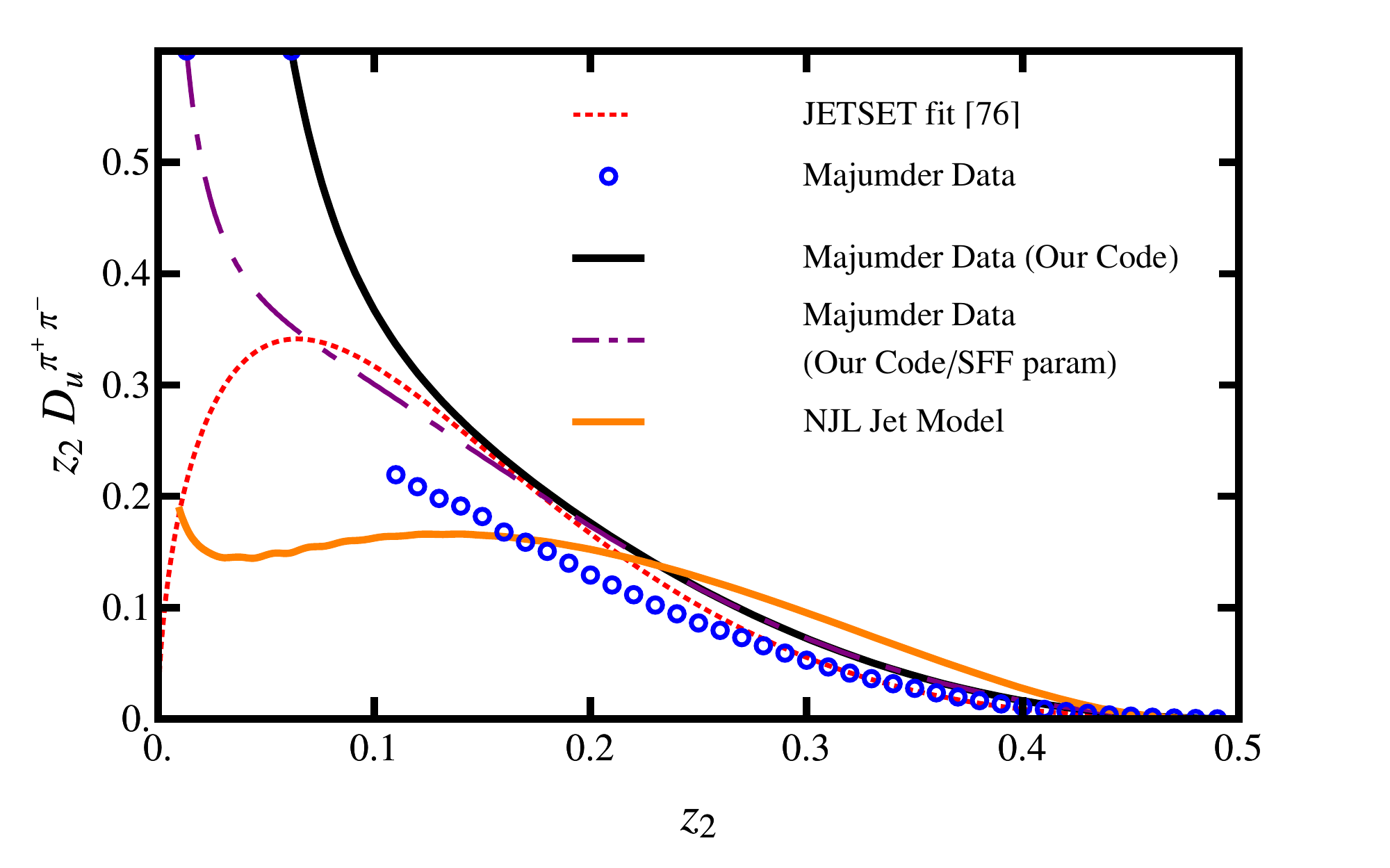}}
\subfigure[$\,z_1=0.5$: $z_2\,D^{\pi^+ \pi^-}_g$]{\label{fig:comparegpppm}
\includegraphics[width=0.48\textwidth]{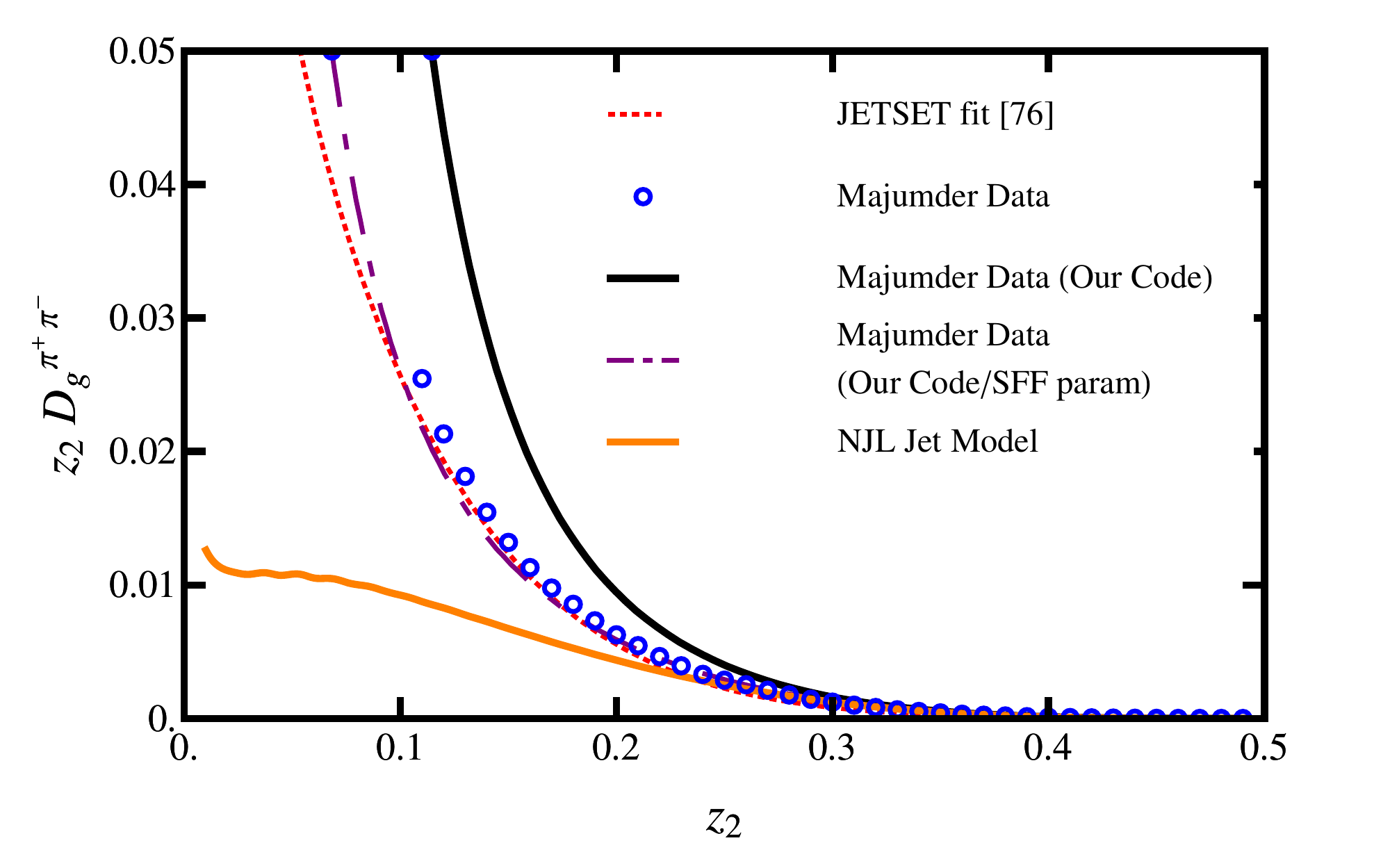}}
\caption{$\pi^+ \pi^-$ dihadron fragmentation functions for $z_1=0.5$ at $Q^2 =109 \text{ GeV}^2$ for a fragmenting ~\subref{fig:compareupppm} $u$ quark and ~\subref{fig:comparegpppm} gluon - see text for details.}
\label{fig:compare1}
\end{center}
\end{figure}

Using the initial parameterized DFFs at $Q^2=2 \mathrm{~GeV}^2$, in Figs.~\ref{fig:compare1} we present the comparison of the parameterized $\pi^+ \pi^-$ up quark and gluon DFFs obtained from JETSET at $Q^2=109\mathrm{~GeV}^2$~(dotted red line) with the evolved solutions of Ref.~\cite{Majumder:2004br}~(blue circles). The solutions obtained using our code on the same initial parameterized DFFs~(solid black line) and the solution to NJL-jet model DFFs evolved to the same momentum scale~(solid orange line) are shown too\footnote{These comparisons are at best semi-quantitative as we do not know the value of $\Lambda_{QCD}$ used in Ref.~\cite{Majumder:2004br}.}. We also consider solutions for the parameterized DFFs evolved using an altered version of our code that treats the QCD evolution of the SFFs with the same parameterized evolution as in Ref.~\cite{Majumder:2004br}~(purple dot-dashed line), rather than using the evolution equations. This serves the purpose of exhibiting how well our code reproduces the parameterized JETSET results. 

The results for the NJL-jet model evolved to $Q^2=109\mathrm{~GeV}^2$ are similar to the parameterized JETSET results of Ref.~\cite{Majumder:2004br} for values of $z_2$ above $0.2$ for both the up quark~(Fig.~\ref{fig:compareupppm}) and gluon~(Fig.~\ref{fig:comparegpppm}) DFFs. Below $z_2=0.2$, our solutions are smaller. Such differences may be expected as the parameterization in Ref.~\cite{Majumder:2004br} overestimates the JETSET results in the low $z_1$ and $z_2$ regions, and so the NJL-jet model results may well be closer to the actual JETSET output. 

We also observe that for the up quark DFF~(Fig.~\ref{fig:compareupppm}), the solution for the parameterized JETSET input evolved using our code produces similar results to the parameterized solution of the JETSET results at $Q^2=109\mathrm{~GeV}^2$ for values of $z_2$ greater than approximately $0.1$. The gluon DFF~(Fig.~\ref{fig:comparegpppm}) solutions differ only at values of $z_2$ lower than $0.25$. In order to understand this difference we explored using the parameterized evolution of the SFFs~\cite{PhysRevD.52.4947} used by Ref.~\cite{Majumder:2004br}. This produced an improved comparison between the parameterized JETSET solution and the DFFs obtained through our code. It is shown that by employing the parameterized SFF evolution we produce results that are similar to the parameterized JETSET solutions for both the up quark~(Fig.~\ref{fig:compareupppm}) and gluon~(Fig.~\ref{fig:comparegpppm}) for values of $z_2$ above approximately $0.1$. 

\begin{figure}[H!t]
\begin{center}
\subfigure[$\,z_1=0.5$: $z_2\,D^{\pi^+ \pi^-}_u$]{\label{fig:evolveupppm}
\includegraphics[width=0.48\textwidth]{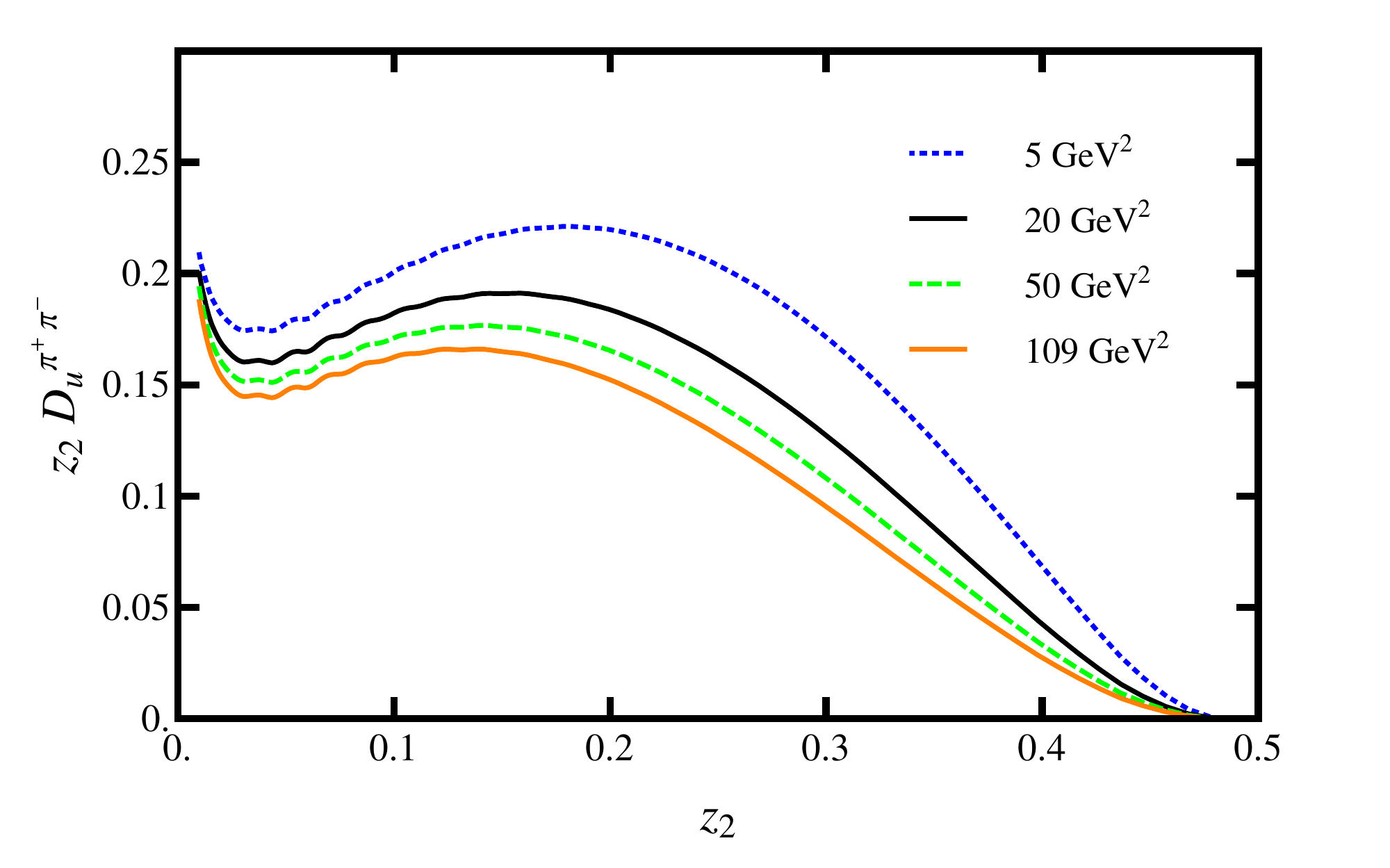}}
\subfigure[$\,z_1=0.5$: $z_2\,D^{\pi^+ \pi^-}_g$]{\label{fig:evolvegpppm}
\includegraphics[width=0.48\textwidth]{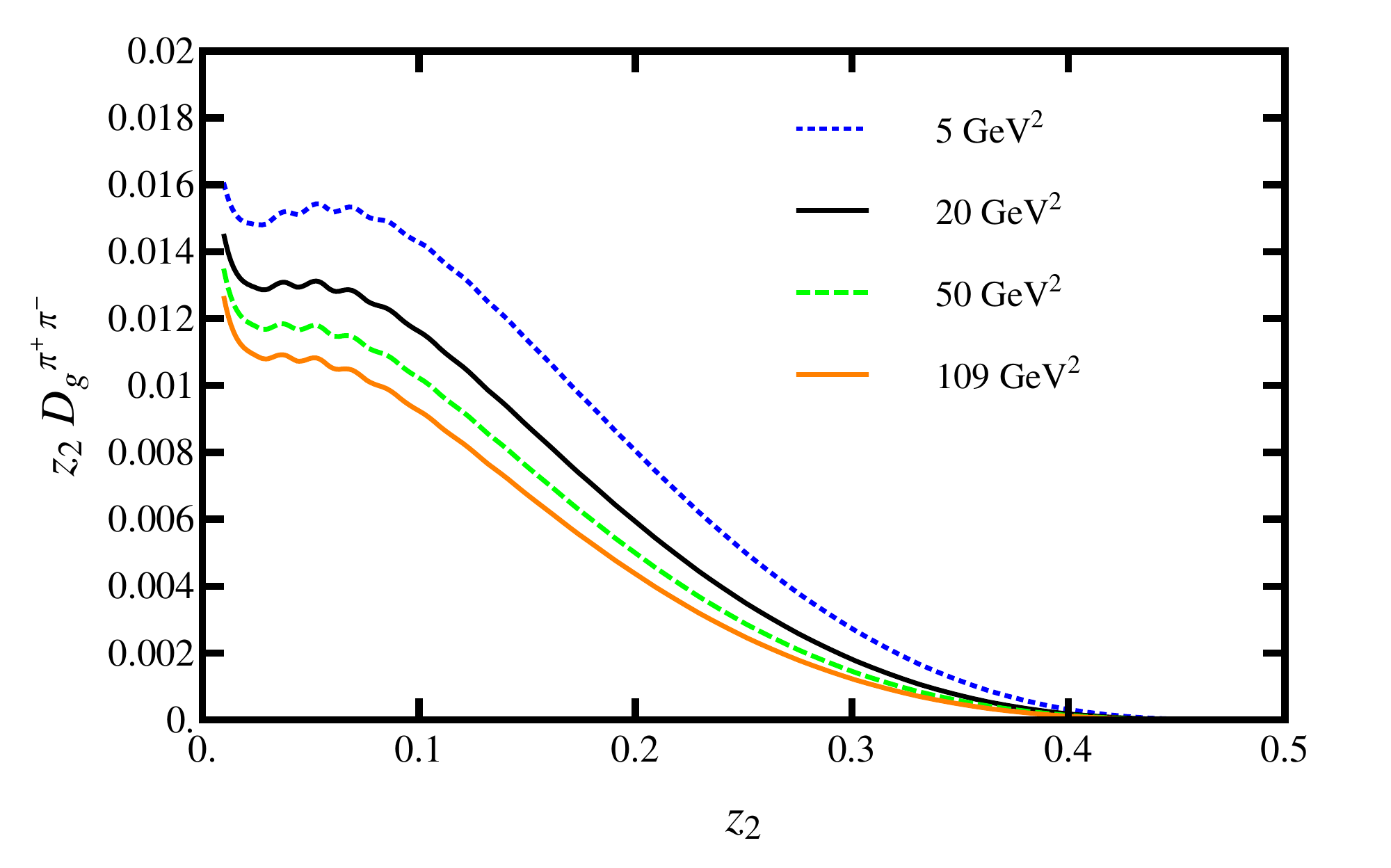}}
\caption{$\pi^+ \pi^-$ dihadron fragmentation functions for the NJL-jet model, for $z_1=0.5$ at $Q^2 =5\text{ GeV}^2 ,~20\text{ GeV}^2,~50\text{ GeV}^2\text{ and}~109 \text{ GeV}^2$ for a fragmenting ~\subref{fig:evolveupppm} $u$ quark and ~\subref{fig:evolvegpppm} gluon.}
\label{fig:evolve1}
\end{center}
\end{figure}

In Figs.~\ref{fig:evolve1}, we present the results for the NJL-jet model DFFs evolved to a range of $Q^2$ values: $5\text{~GeV}^2$~(blue dotted line), $20\text{~GeV}^2$~(black solid line), $50\text{~GeV}^2$~(green dashed line) and $109\text{~GeV}^2$~(orange solid line). For $z_1=0.5$, the results show that as $Q^2$ increases, the DFFs appear to gradually reduce. The peak value is also observed to shift towards lower $z_2$ values.

\section{Conclusions and Outlooks}

In this article, solutions are presented for dihadron fragmentation functions from the NJL-jet model evolved to a typical experimental scale of $Q^2=4~\mathrm{GeV}^2$, from the model scale of $Q^2_0=0.2~\mathrm{GeV}^2$. We first presented a brief summary of the integral equations used to obtain the model scale SFFs and DFFs in Section~\ref{sec:qsffdff}. Sections~\ref{sec:evolsff} and~\ref{sec:evoldffs} describe the numerical method used to solve the evolution equations for SFFs and DFFs, respectively. The QCD evolution equations for the SFFs and the Fortran code used to solve them was based on the method described in Refs.~\cite{Miyama:1995bd,Hirai:1996hv,Hirai:1997gb,Hirai:2011si}. The method used rearranges the evolution equations into non-singlet quark and coupled plus-type quark and gluon equations, followed by discretizing the variables $z$ and $t$ and converting the integral terms in to sums over the integration variable. The same method is employed to solve the QCD evolution equations for the DFFs with the variables $z_1$, $z_2$ and $t$ being discretized.

Section~\ref{sec:evolresults} compares the model scale DFFs with the evolved DFFs for $\pi^+ \pi^-$, $\pi^+ K^-$ and $K^+ K^-$. In Section~\ref{sec:upplpmievol} we investigated the evolution of $D^{\pi^+ \pi^-}_u$ by comparing the model scale and evolved scale DFFs when either $z_1$~(Fig.~\ref{fig:upplpmiz105}) or $z_2$~(Fig.~\ref{fig:upplpmiz205}) is equal to $0.5$. We also considered $z_1$~(Fig.~\ref{fig:upplpmiz102}) or $z_2$~(Fig.~\ref{fig:upplpmiz202}) equal to $0.2$. For the fragmentation of the up quark to $\pi^+ \pi^-$ we noted that the up quark was a favored emission channel for the $\pi^+$, while the down quark produced after the up quark fragments to a $\pi^+$ is a favored emission channel for the $\pi^-$. The evolved DFF showed a shift in the peak value towards the lower $z$ regions, with each plot showing the evolved DFF obtaining a larger magnitude than the model scale DFF in the lower $z$ region.

The focus of Section~\ref{sec:upplkmievol} is on the evolution of $D^{\pi^+ K^-}_u$. Similar to $D^{\pi^+ \pi^-}_u$, the up quark is the favored emission channel for $\pi^+$, however the produced down quark is an unfavored emission channel for $K^-$. The magnitude of the model scale DFF was significantly smaller for $D^{\pi^+ K^-}_u$~(Figs.~\ref{fig:upplkmi}) than for $D^{\pi^+ \pi^-}_u$~(Figs.~\ref{fig:upplpmi}) at light-cone momentum fractions fixed to $0.5$. After evolution, $D^{\pi^+ K^-}_u$ displayed a similar shift in the peak value towards the low $z$ region. For $z_1=0.5$~(Fig.~\ref{fig:upplkmiz105}), the evolved DFF does not obtain a larger magnitude than the model scale DFF in the lower $z_2$ region, whereas for $z_2=0.5$~(Fig.~\ref{fig:upplkmiz205}) the evolved DFF obtains a substantial increase over the model scale DFF in the lower $z_1$ region. This demonstrates the effect evolution has on favored and unfavored emission channels.

Finally, Section~\ref{sec:qkplkmievol} demonstrates the evolution of $D^{K^+ K^-}_q$ for $q=u$, $d$ or $s$. $D^{K^+ K^-}_q$ has favored fragmentation channels for both the up quark and strange quark. This is observed in the results presented in Figs.~\ref{fig:kplkmi} where both $D^{K^+ K^-}_u$ and $D^{K^+ K^-}_s$ have large peaks in the upper $z_2$ region. Both $D^{K^+ K^-}_u$ and $D^{K^+ K^-}_s$ display the shift of the peak value to the lower $z$ region that has been shown in other favored emission channels at light-cone momentum fractions of $0.5$. The down quark is an unfavored emission for both $K^+$ and $K^-$, and so $D^{K^+ K^-}_d$ has a very small magnitude at the model scale. Evolving $D^{K^+ K^-}_d$ shows a considerable increase in the lower $z_2$ region, though the magnitude is still much lower than that of $D^{K^+ K^-}_u$ and $D^{K^+ K^-}_s$ at $z_1=0.5$.

Evolution of the DFFs has the effect of reducing the magnitudes at higher $z$, resulting in peaks occuring earlier in the range of $z$ values with a reduced magnitude. If the magnitude of the DFF was small at the model scale, a significant increase in the magnitude at the low $z$ region is observed after evolution. The first of these two effects generally occurs for the favored emission channels, where the light-cone momentum fraction of the emitted hadron is not in the low $z$ region. The second effect, typically occurs when the fragmentation channel is unfavored or when the emitted hadron carries a small light-cone momentum fraction. 

In Section~\ref{sec:comparison}, we evolve the parameterized JETSET data at $2\mathrm{~GeV}^2$ from Ref.~\cite{Majumder:2004br} to $109\mathrm{~GeV}^2$ using our code to compare the solutions obtained with the parameterized JETSET data at the same scale for both the up quark and gluon fragmenting to $\pi^+ \pi^-$~(Figs.~\ref{fig:compare1}). We also presented solutions for the NJL-jet model up quark and gluon DFFs evolved to $Q^2$ values of $5\text{~GeV}^2$, $20\text{~GeV}^2$, $50\text{~GeV}^2$ and $109\text{~GeV}^2$~(Figs.~\ref{fig:evolve1}). The solutions show that for $z_1=0.5$, the DFFs are reduced as $Q^2$ increases and the peak value shifts towards the lower $z_2$ region.

Extensions of the NJL-jet model for single hadron fragmentation functions such as the inclusion of hadronic resonances and their decays~\cite{Matevosyan:2011ey} and inclusion of the transverse momentum dependence~\cite{Matevosyan:2011vj} have been accomplished using a Monte Carlo framework. These extensions are possible for DFFs as well, but they are beyond the scope of this work are left for the future.

\acknowledgements{
This work was supported by the Australian Research Council through Australian Laureate Fellowship FL0992247~(AWT), the ARC Centre of Excellence for Particle Physics at the Terascale and by the University of Adelaide.}

\bibliographystyle{apsrev}
\bibliography{dihadevolv14arXiv}

\end{document}